\documentclass[12pt]{article}
\usepackage{cite}
\usepackage{amsmath, amsthm, amssymb, graphicx,slashed}

\numberwithin{equation}{section}

\def\p{\partial}

\def\be{\begin{equation}}
\def\ee{\end{equation}}
\def\ba{\begin{align}}
\def\ea{\end{align}}
\def\beq{\begin{eqnarray}}
\def\eeq{\end{eqnarray}}

\def\a{\alpha}
\def\b{\beta}

\begin{document}

\begin{titlepage}

\vskip 1.5in
\begin{center}
{\bf\Large{Elliptic genera and real Jacobi forms} }\vskip 1.9cm
{ Sujay K. Ashok${}^a$ and Jan Troost${}^b$} \vskip
0.3in

 \emph{$^{a}$}
\emph{ Institute of Mathematical Sciences \\
   C.I.T Campus, Taramani\\
   Chennai, India 600113\\ 
}


  \emph{\\${}^{b}$ Laboratoire de Physique Th\'eorique}\footnote{Unit\'e Mixte du CNRS et
     de l'Ecole Normale Sup\'erieure associ\'ee \`a l'universit\'e Pierre et
     Marie Curie 6, UMR
     8549.
}  \\
 \emph{ Ecole Normale Sup\'erieure  \\
 24 rue Lhomond \\ F--75231 Paris Cedex 05, France}
\end{center}
 \vskip 0.5in

\baselineskip 16pt

\begin{abstract}

  \vspace{.2in} 

  We construct real 
Jacobi forms with matrix index using path
  integrals. The path integral expressions represent elliptic genera
  of two-dimensional ${\cal N}=(2,2)$ supersymmetric theories. They arise in
  a family labeled by two integers $N$ and $k$ which determine the
  central charge of the infrared fixed point through the formula $c= 3
  N ( 1+ 2N/k)$.  We decompose the real Jacobi form into a mock
  modular form and a term arising from the continuous spectrum of the
  conformal field theory. For a given $N$ and $k$ we argue that the
  Jacobi form represents the elliptic genus of a theory defined on 
a $2N$ dimensional linear dilaton background with
  $U(N)$ isometry, an asymptotic circle of radius $\sqrt{k \alpha'}$
  and linear dilaton slope $N \sqrt{2/k}$.  We also present formulas
  for the elliptic genera of their orbifolds.

\end{abstract}
\end{titlepage}
\vfill\eject

\tableofcontents 


\section{Introduction}
Despite a long and rich history, quantum field theory continues to
develop in surprising ways. In particular, we continue to discover new
classes of conformal field theories in various dimensions with novel
properties. While we do not always have a handle on the full spectrum
of the conformal field theory, or even a Lagrangian description, we
can often give a catalog of detailed properties of the conformal
field theory through other means.

In this paper, we concentrate on conformal field theories in two
dimensions with ${\cal N}=(2,2)$ supersymmetry. For these theories, we have
characterizing properties like the central charge, the Witten index,
the spectrum of chiral primaries, the elliptic genus, 
three-point functions, boundaries preserving
conformal symmetry 
etc. Often we are only able to specify part
of the characterizing properties of the conformal field theory.  In
this paper we propose path integral expressions for the elliptic genus
of certain conformal field theories, and we give good evidence for an
identification of the conformal field theories in question.

The elliptic genus is a weighted
trace that captures short representations in the spectrum
\cite{Schellekens:1986yi,Witten:1986bf,Eguchi:1988vra,Kawai:1993jk,Witten:1993jg}, as well as
rough characteristics of the continuum \cite{Troost:2010ud}.
In the Hamiltonian formalism, it is given by a trace
over the Hilbert space
\be
\chi(\tau, \a)= \text{Tr} 
(-1)^F\,  z^{J_0} q^{L_0- \frac{c}{24}} \bar{q}^{\bar{L}_0- \frac{c}{24}} \,.
\ee
Here, $q=e^{2\pi i \tau}$ and $z=e^{2\pi i \a}$. 
The scaling operator 
$L_0$ measures the conformal dimension of left movers and the operator $J_0$
measures the left moving $R$-charge. In cases 
for which the spectrum is discrete, 
the insertion $(-1)^F$ projects onto the
right moving ground states and the resulting elliptic genera are holomorphic Jacobi forms,
i.e. they have well defined modular and elliptic properties and
are holomorphic functions of $q$ and $z$.
However,  when the conformal field theory in
question has a continuum of states the elliptic genus is typically not
holomorphic \cite{Troost:2010ud}. It is the modular completion of a mock Jacobi
form \cite{Zwegers,Zagier}. The mock Jacobi form is holomorphic and arises from 
 right-moving ground state contributions.
The modular completion arises from the continuum, which contributes even though the 
right-movers are not in the ground state. The resulting expression is a real Jacobi form.

The analysis of \cite{Troost:2010ud}  was generalized in \cite{Eguchi:2010cb, Ashok:2011cy} in which a
twisted elliptic genus of the supersymmetric coset
$SL(2,\mathbb{R})/U(1)$ or cigar conformal field theory was calculated
from the path integral formalism. It was made clear in
\cite{Troost:2010ud,Ashok:2011cy,Ashok:2013kk} that the non-holomorphic contribution arises due to
a mismatch in the spectral density of bosons and fermions in the
continuum sector. The mismatch can be characterized in terms of the 
asymptotic expression for the supercharge.

In this article we propose path integral expressions for twisted
elliptic genera that are 
generalizations of the path integral expression for
the cigar and Liouville conformal field theory \cite{Troost:2010ud,Ashok:2011cy}. 
The central charge of the conformal field theories is 
$c=3 N (1 + 2N/k)$. They are further characterized by ${\cal N}=2$ superconformal symmetry and 
at least a $U(1)^N$ global symmetry.
 In mathematical terms, our
expression for the elliptic genus is described as the modular
completion of a mock Jacobi form with matrix index. The Jacobi forms
depend on a chemical potential for $R$-charge and $N$ chemical potentials
for the global symmetry group $U(1)^N$.

Finally, we argue for the identification of the conformal field theory that gives rise
to these elliptic genera. The generalization of the cigar elliptic genus is the elliptic
genus of linear dilaton spaces of dimension $2N$, discovered in \cite{Kiritsis:1993pb} and
further analyzed in \cite{Hori:2001ax,Hori:2002cd} as well as \cite{Ashok:2013kk}. In particular,
we are able to provide a detailed identification of contributions to the elliptic genus
by the wound, bound strings of \cite{Ashok:2013kk}.

This paper is organized as follows. In section \ref{pathprop}, we
briefly review relevant properties of the elliptic genus of the cigar
conformal field theory.  We propose our generalization to a 
class of models labeled by two integers $k$ and $N$. The proposal has
good modular and elliptic properties, namely it is a real
Jacobi form with matrix index. In section \ref{decomposition} we
derive the holomorphic component of the elliptic genus, and provide
various forms of it and a detailed decomposition in terms of
characters.  We also identify the covariantization provided by the
path integral expression.  In section \ref{KKL}, we compare the
properties of the proposed elliptic genus with properties of 
asymptotic linear dilaton models and heuristically connect to a gauged linear
sigma model description of these models. 
In section \ref{genLiou}, we propose path integral expressions for
generalizations of ${\cal N}=2$ Liouville theories and show that a
diagonal orbifold leads back to the generalized cigar elliptic genus.  
We conclude in section \ref{conclusions}.

\section{A path integral elliptic genus}
\label{pathprop}
In this section, we review the path integral result for the elliptic genus of the ${\cal N}=2$ superconformal 
cigar conformal field theory \cite{Troost:2010ud,Eguchi:2010cb, Ashok:2011cy}. 
We then generalize the path integral expression to
a proposal for
the elliptic genus for a two-dimensional ${\cal N}=2$ superconformal
field theory in two dimensions labeled by an extra integer $N$ and
with at least a $U(1)^N$ global symmetry. We show that our proposal is
a real Jacobi form of matrix index. It is a function of a modular
parameter $q=e^{2 \pi i \tau}$, an $R$-charge fugacity $z=e^{2 \pi i
  \alpha}$ as well as $N$ chemical potentials $y_i = e^{2 \pi i
  \beta_i}$. 

\subsection{The cigar elliptic genus}
\label{cigar}
The path integral calculation of the cigar elliptic genus gives rise to the result: 
\be
\label{cigarpathintegral}
 \chi_{cos}(\tau,\a) = k \int_0^1 ds_1
ds_2 \sum_{m,w \in \mathbb{Z}}\left[\frac{\theta_{11}(\tau,s_1\tau+s_2-\a-\frac{\a}{k}
    )}{\theta_{11}(\tau,s_1\tau+s_2-\frac{\a}{k})}\right]\,
e^{2\pi i \a w}\, e^{-\frac{\pi k}{\tau_2}|m+w \tau +(s_1\tau+s_2
  )|^2}\,.  
\ee
 The calculation of the path integral in the end  boils down to the contribution of
elementary ingredients \cite{Ashok:2011cy}.
There is the contribution of the zero modes of 
a charged complex boson of radius $\sqrt{k \alpha'}$,
a charged complex fermion of charge one, a complex boson of charge one, and Wilson lines
$s_1$ and $s_2$ for the corresponding $U(1)$ gauge group.
The complex boson has $R$-charge $1+1/k$ while the complex fermion has $R$-charge $1/k$.
The phase $e^{2 \pi i \alpha w}$ ensures periodicity in the Wilson line $s_1$ and can be argued
on the basis of modular covariance or a chiral fermion anomaly.
These ingredients lead to the final result (\ref{cigarpathintegral}).

\subsection{A proposed path integral generalization}
\label{proposal}

We propose the following generalized path integral result:
\be\label{Nguess}
\chi_{{N}}(\tau,\a,\b_i) = k
\int_{0}^1 ds_1 ds_2 \sum_{m,w \in \mathbb{Z}} \prod_{i=1}^N \left[\frac{\theta_{11}(\tau,s_1\tau+s_2-\a-\frac{N\a}{k}+\beta_i)}{\theta_{11}(\tau,s_1\tau+s_2-\frac{N\a}{k}+\beta_i)}\right]\, 
e^{2\pi i N \a w}\, 
e^{-\frac{\pi k}{\tau_2}|m+w \tau +(s_1\tau+s_2 )|^2}\, ,
\ee
where the case $N=1$ correspond to the cigar elliptic genus (\ref{cigarpathintegral}).
We can identify the final ingredients that make up this path integral.
We have a
model with $U(1)$ gauge field under which we have $N$ charged complex
fermions and bosons, each carrying charge $1$. We also have the zero mode of a charged
boson of radius $\sqrt{k \alpha'}$. The
 complex bosons have $R$-charge $1+N/k$, and the $R$-charge of the fermions
differs from these by one. The anomalous phase factor again ensures periodicity in the 
holonomy $s_1$ of the $U(1)$ gauge field. Each complex boson and fermion is charged under
a global $U(1)^N$ symmetry. 

Another useful perspective on the path integral arises when we rewrite the expression in terms of holonomies taking values on the real line:
\be
\chi_{{N}}(\tau,\a,\b_i) = k
\int_{-\infty}^{+\infty} ds_1 ds_2  \prod_{i=1}^N 
\left[\frac{\theta_{11}(\tau,s_1\tau+s_2-\a-\frac{N\a}{k}+\beta_i)}{\theta_{11}(\tau,s_1\tau+s_2-\frac{N\a}{k}+\beta_i)}\right]\, 
e^{-\frac{\pi k}{\tau_2}|s_1\tau+s_2|^2}\, .
\ee
We note that near a multiple pole of order $m$ of the ratio of $\theta_{11}$ functions,
we can write an approximation to the integral in terms of a coordinate $z$ near the
multiple pole as:
\begin{eqnarray}
\int dz d \bar{z} 
\frac{1}{z^m} \sum_{n,\bar{n} \ge 0} c_{n,\bar{n}} z^n \bar{z}^{\bar{n}} \, .
\end{eqnarray}
The angular integration will give zero unless $n-m=\bar{n} \ge 0$, and therefore
our integral expression is free of divergences.

In summary, we propose that this expression is the result of computing a
generalized elliptic genus of a two-dimensional conformal field theory
with ${\cal N}=2$ superconformal symmetry and at least a $U(1)^N$ global symmetry
through path integral means. We give evidence for what this theory
is in section \ref{KKL}. In this section we take equation \eqref{Nguess}
as our starting point and analyze its modular and elliptic
properties.

\subsection{A real Jacobi form of matrix index}

After double Poisson resummation, we obtain an expression with which it is easier
to check the modular properties:
\begin{multline}\label{modular}
\chi_{{N}}(\tau,\a,\b_i) = \int_0^1 ds_1 ds_2 \sum_{m,w \in \mathbb{Z}}\prod_{i=1}^{N}\left[\frac{\theta_{11}(\tau,s_1\tau+s_2-\a-\frac{N\a}{k}+\b_i)}{\theta_{11}(\tau,s_1\tau+s_2-\frac{N\a}{k}+\b_i)}\right]\cr
\times e^{-2\pi i s_2 w}e^{2\pi i s_1(m-N\a)}
e^{-\frac{\pi}{k\tau_2}|m-N\a+w\tau|^2}\,.
\end{multline}
 We consider the behavior of
the elliptic genus under the action of the modular group on
$\tau$. Since this operation interchanges the cycles of the torus, it
must be accompanied by the corresponding transformation of the
holonomies and the winding numbers. One can check that
the following operation leaves the expression invariant:
\begin{align}
\tau \rightarrow \tau+1 \qquad m\rightarrow m-w \qquad s_2\rightarrow s_2 -s_1 \,.
\end{align}
The measure factor does not change and we used that the integrand is periodic in the
holonomies.
This shows invariance under the $T$ transformation. In order to obtain the transformation of the elliptic genus under the $S$-transformation
\begin{align}
\tau &\rightarrow -\frac{1}{\tau}\qquad \a \rightarrow \frac{\a}{\tau} \qquad \beta_i \rightarrow \frac{\beta_i}{\tau} \cr
s_1 & \rightarrow -s_2 \qquad s_2\rightarrow s_1  \qquad
w \rightarrow -m \qquad m \rightarrow w \, ,
\end{align}
we use the transformation of the Jacobi theta function
\be
\theta_{11}(-\frac{1}{\tau}, \frac{\a}{\tau}) = -i(-i\tau)^{\frac{1}{2}}\, e^{\frac{\pi i \a^2}{\tau}}\theta_{11}(\tau, \a) \,.
\ee
The phase factor that one picks up is given by:
\begin{align}
 &\frac{\pi i}{\tau} \sum_{i=1}^N \left( -2\a(s_1\tau+s_2-\frac{N\a}{k}+\beta_i) 
+ \a^2 \right)+\frac{2\pi iN\a}{\tau}(s_1\tau+s_2) 
\cr
 &= \frac{\pi i \alpha^2}{\tau} \frac{c}{3} - \frac{\pi i}{\tau} 2 \alpha \sum_{i=1}^N \beta_i 
\, ,
\end{align}
where we  introduced the central charge parameter:
\be
c = 3 N ( 1 + \frac{2N}{k}) \, .
\ee
At this stage, we will suppose that the quantity:
\be
M = \frac{k}{N} \, ,
\ee
is an integer, although this is not strictly necessary. If it is, our expression
is elliptic in $\alpha$ where we can perform shifts by multiples of $M$.\footnote{If $M$ were not
integer, we would continue the discussion in terms of shifts by multiples of
 $k$.} The parameters $\beta_i$ are elliptic in integer multiples of
the periods of the torus.
 We
summarize the modular and elliptic properties:
\begin{align}\label{cigarmod-ell}
\chi_{ {N}}(\tau+1,\a,\b_i) &=\chi_{ {N}}(\tau,\a,\b_i) \cr
\chi_{ {N}}(-\frac{1}{\tau},\frac{\a}{\tau},\frac{\b_i}{\tau}) &= 
e^{\frac{\pi i\frac{c}{3} \a^2}{\tau}} e^{-\frac{\pi i}{\tau}2 \alpha \sum_i \b_i}\chi_{ {N}}(\tau,\a,\b_i) \quad \text{where all $\b_i$ are rescaled.}\cr
\chi_{ {N}}(\tau,\a+M
,\b_i) &=(-1)^{k}\chi_{ {N}}(\tau,\a,\b_i)\cr
\chi_{ {N}}(\tau,\a+M 
 \tau,\b_i) &=(-1)^{k} 
e^{-\pi i \frac{c}{3} (M^2 \tau + 2 M \a)} e^{2\pi i M \sum_{i}\b_i}\chi_{ {N}}(\tau,\a,\b_i) \cr
\chi_{ {N}}(\tau,\a,\b_i+1) &=\chi_{ {N}}(\tau,\a,\b_i)\cr
\chi_{ {N}}(\tau,\a,\b_i+\tau) &=e^{2\pi i \a}\chi_{ {N}}(\tau,\a,\b_i) \quad \text{where only a single $\b_i$ is shifted}\,.
\end{align}
We have a real Jacobi form with matrix index  (see e.g. \cite{Skoruppaetal}) given by
\be
\left( \begin{array}{cccc}
 -\frac{c}{3} & 1 & 1 & \dots 
\\
 1 & 0 & 0 & \dots
\\ 1 & 0 & 0 & \dots
\\
\dots & \dots & \dots & \dots 
\end{array}
\right) \, .
\ee
We have made allowance for a parameter $\alpha$ that is normalized in accordance with standard
physics conventions. If one wants to renormalize $\alpha$ to have periods $(1,\tau)$, one should rescale
the entries in the table accordingly. Finally, we note the charge conjugation symmetry:
\be\label{cc}
\chi_N(\tau,-\alpha,-\beta_i) = \chi_N(\tau,\alpha,\beta_i)
\, .
\ee
In summary, we have a path integral expression for a real Jacobi form
with matrix index. The
Lagrangian perspective renders manifest the elliptic and modular properties of the elliptic genus.

\section{The mock modular form}
\label{decomposition}
In this section, we would like to find a Hamiltonian interpretation of the path integral expression that
we proposed in section  \ref{pathprop}. We wish to distinguish holomorphic contributions,  arising
from right-moving ground states in the underlying conformal field theory, and a remainder term that originates
in the continuous part of the spectrum. We present an interpretation of the holomorphic contribution in terms
of ${\cal N}=2$ superconformal algebra characters, and offer a rewriting in terms of a
contour integral. The latter is reminiscent of the expressions for 
purely holomorphic elliptic genera arising from gauged linear sigma-models.

\subsection{The Hamiltonian viewpoint}
In the first few steps, we prepare the ground for an interpretation of our expression in terms of a physical
state sum by going to Hamiltonian variables and executing the integral over the holonomies.
We start by singly Poisson resumming our proposal to find after relabeling:
\be
\chi_{ N} (\tau,\a,\b_i) = \sqrt{k \tau_2} \int_0^1 ds_{1,2} \sum_{n,w\in \mathbb{Z}} \prod_{i=1}^N
\left[\frac{\theta_{11}(\tau,s_1\tau+s_2-\a-\frac{N\a}{k} +\beta_i)}{
\theta_{11}(\tau,s_1\tau+s_2-\frac{N\a}{k}+\beta_i)}\right]\, 
e^{2\pi i N\a w}\, 
e^{-2\pi i s_2 n}\,
q^{\ell_0} \bar{q}^{\bar{\ell}_0} \, ,
\ee
where we introduced contributions $\ell_0$ and $\bar{\ell}_0$ to the conformal dimensions equal to
\be\label{ellandellbar}
\ell_0 = \frac{(n-k(w+s_1))^2}{4k} \qquad \bar{\ell}_0 = \frac{(n+k(w+s_1))^2}{4k} \, .
\ee
These correspond to momenta and winding of a boson of radius $\sqrt{k \alpha'}$, twisted by 
the holonomies $s_{1,2}$ of the $U(1)$ gauge field.
%
%
We expand the denominator, using the formula
\be\label{denomexpand}
\frac{1}{i \theta_{11}(\tau, \a)} = \frac{1}{\eta^3(\tau)}  \sum_{r \in \mathbb{Z}}z^{r+\frac{1}{2}}S_{r}(q) \,,
\ee
%
where $q=e^{2\pi i \tau}$ and $z=e^{2\pi i \a}$ and
\be
S_r(q) = \sum_{n=0}^{\infty}(-1)^nq^{\frac{n(n+2r+1)}{2}}\,.
\ee
In the expression for the elliptic genus, we use this expansion for each theta function in the denominator; for each such term, the  argument $z$ is given by
\be
x_i=q^{s_1}e^{2\pi is_2}z^{-\frac{N}{k}} y_i\,,
\ee
%
where $y_i = e^{2\pi i \b_i}$. 
As proven in appendix \ref{idproof}, the expansion is valid when $|q|<|x_i|<1$ which we will assume from now on. Note in particular
that when $|y_i|=|z|=1$, we have that $|q|<|q^{s_1}|<1$ for $s_1$ between $0$ and $1$. Here, the decomposition
of the holonomy plane into periodic variables $s_{1,2}$ and integer parts $(m,w)$ plays a crucial role.
We also write the theta series in the numerator as:
\be\label{numexpand}
\theta_{11} (\tau, \a) = i \sum_{m \in \mathbb{Z}} (-1)^{m}q^{\frac{(m-\frac{1}{2})^2}{2}} \, z^{-m+\frac{1}{2}}
\ee
where, for each term in the product,
 the argument $z$ is given by
\be
z_i = q^{s_1}e^{2\pi is_2}z^{-1-\frac{N}{k}} y_i \, .
\ee
We then collect terms and perform the integral over the holonomy $s_2$ which imposes the constraint
of gauge invariance on the physical state space:
\be
\sum_{i=1}^N (r_i-m_i+1)= n \,.
\ee
The constraint leads to simplifications that give rise to:
\begin{multline}
\chi_{ N} (\tau,\a,\b_i)=  (-1)^N \frac{\sqrt{k \tau_2}}{\eta^{3N}(\tau)}\sum_{m_i,r_i,n,w}
\int_0^1 ds_1 (-1)^{\sum_i m_i}q^{\frac{1}{2}\sum_i (m_i-\frac{1}{2})^2} \, z^{\sum_i (m_i-\frac{1}{2})}\cr
q^{-nw}z^{Nw-\frac{Nn}{k}}(q\bar{q})^{\bar{\ell}_0} \delta_{\sum_i(r_i-m_i+1) -n}  \prod_{i=1}^N y_i^{r_i-m_i+1} S_{r_i}(q)  \,.
\end{multline}
We will henceforth omit writing the explicit argument of the function
$S_r(q)$ in order to make the formulas less cumbersome.  To linearize
the integration over the second holonomy $s_1$, we introduce the
integration over a variable $s$ that will have an interpretation as a
non-compact (radial) momentum. We moreover introduce the right-moving
momentum $v$ on the circle of radius $\sqrt{k \alpha'}$:
\be
v = n +kw \, .
\ee
We end up with 
\begin{multline}
\chi_{ N}(\tau,\a,\b_i)=(-1)^N\frac{2 \tau_2 }{\eta^{3N}(\tau)}\sum_{m_i,r_i,v,w}\int_{-\infty}^{\infty}ds\int_0^1 ds_1 (-1)^{\sum_i m_i}
q^{\frac{1}{2}\sum (m_i-\frac{1}{2})^2} \, 
z^{\sum_i (m_i-\frac{1}{2})}\cr
q^{kw^2-vw}z^{N(2w-\frac{v}{k})}
(q\bar{q})^{\frac{v^2}{4k} + \frac{s^2}{k} + s_1(is+\frac{v}{2})}
\delta_{\sum_i(r_i-m_i+1) -(v-kw)} \prod_{i=1}^N y_i^{r_i-m_i+1} S_{r_i}  \,.
\end{multline}
Finally, performing the $s_1$ holonomy integral,  we get:
\begin{multline}\label{final}
\chi_{ N}(\tau,\a,\b_i)=(-1)^{N+1}\frac{1}{\pi\eta^{3N}(\tau)}\sum_{m_i,r_i,v,w}\int_{- \infty}^{+\infty} \frac{ds}{2is+v} 
(-1)^{\sum_i m_i}q^{\frac{1}{2}\sum (m_i-\frac{1}{2})^2} \, z^{\sum_i (m_i-\frac{1}{2})}\cr
q^{kw^2-vw}z^{N(2w-\frac{v}{k})} 
(q\bar{q})^{\frac{v^2}{4k} + \frac{s^2}{k}}((q\bar q)^{is+\frac{v}{2}}-1)\,  
\delta_{\sum_i(r_i-m_i+1) -(v-kw)} \prod_{i=1}^N y_i^{r_i-m_i+1} S_{r_i}  \,.
\end{multline}
We have prepared the ground for a state sum interpretation, by
performing the integral over holonomies and introducing a non-compact
radial momentum $s$. However, interpretation is still not
straightforward since the expression exhibits an imaginary exponent of
the modular parameter $q$, which we wish to avoid in
a unitary state space sum.  In what follows we perform a slight
variation of the analysis in \cite{Troost:2010ud, Ashok:2011cy} and
extract both a holomorphic mock modular contribution and a remainder
term from this expression, which exhibit exponents corresponding to real
conformal dimensions.

\subsection{The holomorphic, mock modular form}
In equation \eqref{final} we can distinguish two terms. The first term has an imaginary
exponent and is of the form:
\begin{multline}\label{firstpiece}
\chi_{ N,I}(\tau,\a,\b_i)=
(-1)^{N+1} \frac{1}{\pi\eta^{3N}(\tau)}\sum_{m_i,r_i,v,w}\int_{\mathbb{R}+i\epsilon} \frac{ds}{2is+v} 
(-1)^{\sum_i m_i}q^{\frac{1}{2}\sum (m_i-\frac{1}{2})^2} \, z^{\sum_i (m_i-\frac{1}{2})}\cr
q^{kw^2-vw}z^{N(2w-\frac{v}{k})} 
(q\bar{q})^{\frac{v^2}{4k} + \frac{s^2}{k}} \, (q\bar q)^{is+\frac{v}{2}}\, \delta_{\sum_i(r_i-m_i+1) -(v-kw)}\,  
\prod_{i=1}^N y_i^{(r_i-m_i+1)} S_{r_i} \,,
\end{multline}
while the second piece has real exponents and takes the form
\begin{multline}\label{Npart2}
\chi_{ N,II}(\tau,\a,\b_i)=(-1)^{N}\frac{1}{\pi\eta^{3N}(\tau)}\sum_{m_i,r_i,v,w}\int_{\mathbb{R}+i\epsilon}  \frac{ds}{2is+v} 
(-1)^{\sum_i m_i}q^{\frac{1}{2}\sum (m_i-\frac{1}{2})^2} \, z^{\sum_i (m_i-\frac{1}{2})}\cr
q^{kw^2-vw}z^{N(2w-\frac{v}{k})} 
(q\bar{q})^{\frac{v^2}{4k} + \frac{s^2}{k}}\,   \delta_{\sum_i(r_i-m_i+1) -(v-kw)}
\prod_{i=1}^N y_i^{(r_i-m_i+1)} S_{r_i} \,.
\end{multline}
When separating the two terms, a regularization of the pole at $s=0=v$
is required. It introduces a minor ambiguity (related to the arbitrary
separation between discretuum and continuum at zero radial momentum) of little
consequence in the following. (See e.g.  \cite{Troost:2010ud,
  Ashok:2011cy}.)

Our technique will be to shift the contour of the second piece 
\eqref{Npart2} until
part of it combines well with the first 
term, into a holomorphic
discrete contribution. What is left of the second piece, we can then
move back to the real contour of integration, guaranteeing that all
contributions will have a real exponent.

We start with the second 
term and perform the following shifts in the integration and summation variables:
\be
v\rightarrow v + k\qquad w\rightarrow w+1 \qquad s\rightarrow s+\frac{ik}{2} \,.
\ee
%
%
This part of the elliptic genus can then be written as 
\begin{multline}
\chi_{ N,II}(\tau,\a,\b_i)=\frac{(-1)^{N}}{\pi\eta^{3N}(\tau)}\sum_{m_i,r_i,v,w}\int_{\mathbb{R}+i\epsilon-\frac{ik}{2}} \frac{ds}{2is+v} 
(-1)^{\sum_i m_i}q^{\frac{1}{2}\sum (m_i-\frac{1}{2})^2} \, z^{\sum_i (m_i-\frac{1}{2})}\cr
q^{kw^2-vw}z^{N(2w-\frac{v}{k})} 
(q\bar{q})^{\frac{v^2}{4k} + \frac{s^2}{k}} (q\bar{q})^{is+\frac{v}{2}}z^{N} q^{kw-v}\delta_{\sum_{i}(r_i -m_i+1) -v+kw} \,  
\prod_{i=1}^N y_i^{(r_i-m_i+1)} S_{r_i}  \,.
\end{multline}
The $z^N$ factor can be absorbed into each of the $z^{m_i-\frac{1}{2}}$ factors to get $z^{m_i+\frac{1}{2}}$; in order to match this with the expression in part $I$, we define 
\begin{align}
n_i = m_i+1 \qquad \text{ and} \qquad t_i = r_i+1\,.
\end{align}
We also use the identity $S_{t_i-1}=1-S_{-t_i}$ to obtain the following expression:
\begin{multline}\label{chiNIIfinal}
\chi_{ N,II}(\tau,\a,\b_i)=\frac{1}{\pi \eta^{3N}(\tau)}\sum_{n_i,t_i,v,w}\int_{\mathbb{R}+i\epsilon-\frac{ik}{2}} \frac{ds}{2is+v} 
(-1)^{\sum_i n_i}q^{\frac{1}{2}\sum (n_i-\frac{1}{2})^2} \, z^{\sum_i (n_i-\frac{1}{2})}\cr
q^{kw^2-vw}z^{N(2w-\frac{v}{k})} 
q^{\sum_{i}(1-n_i) + kw-v}(q\bar{q})^{\frac{v^2}{4k} + \frac{s^2}{k} + is+\frac{v}{2}}\  
\delta_{\sum_i(t_i-n_i+1)-(v-kw)} \cr
(1-S_{\sum_i (1-n_i) + \sum_{j=1}^{N-1}t_j -v+kw}) \prod_{j=1}^{N-1} (1-S_{-t_j}) 
\prod_{i=1}^Ny_i^{(t_i-n_i+1)}\, . 
\end{multline}
There are $2^N$ terms in the product of $N$ factors, each of the form $(1-S)$. We focus on the monomial arising from the product of $N$ $S$'s. Using the fact that the $q$-exponent can be absorbed by using the formula
\be
q^{r}S_{r} = S_{-r} \, ,
\ee
we obtain the following expression (after relabeling $n_i$ as $m_i$ and $t_i$ as $r_i$):
\begin{multline}\label{chiNIISS}
\chi_{ N,II}^{S}(\tau,\a,\b_i)=\frac{(-1)^{N}}{\pi \eta^{3N}(\tau)}\sum_{m_i,r_i,v,w}\int_{\mathbb{R}+i\epsilon-\frac{ik}{2}} \frac{ds}{2is+v} 
(-1)^{\sum_i m_i}q^{\frac{1}{2}\sum (m_i-\frac{1}{2})^2} \, z^{\sum_i (m_i-\frac{1}{2})}\cr
q^{kw^2-vw}z^{N(2w-\frac{v}{k})} 
(q\bar{q})^{\frac{v^2}{4k} + \frac{s^2}{k}+is+\frac{v}{2}} \,  \delta_{\sum_{i=1}^{N}(r_i -m_i+1) -(v-kw)} \, 
\prod_{i=1}^{N} y_i^{(r_i-m_i+1)} S_{r_i}\,.
\end{multline}
We observe that the integrand above has the same form as that of $\chi_{ N,I}$ in 
equation \eqref{firstpiece}, the difference being the shifted contour of integration. Combining these two terms we obtain the following contour integral:
\begin{multline}
\chi_{ N,hol}(\tau,\a,\b_i)=\frac{(-1)^{N+1}}{\pi \eta^{3N}(\tau)}\sum_{m_i,r_i,v,w}\left[\int_{\mathbb{R}+i\epsilon}-\int_{\mathbb{R}+i\epsilon-\frac{ik}{2}}\right] \frac{ds}{2is+v} 
(-1)^{\sum_i m_i}q^{\frac{1}{2}\sum (m_i-\frac{1}{2})^2} \, z^{\sum_i (m_i-\frac{1}{2})}\cr
(q\bar{q})^{is+\frac{v}{2}} q^{kw^2-vw}z^{N(2w-\frac{v}{k})} 
(q\bar{q})^{\frac{v^2}{4k} + \frac{s^2}{k}} \, \delta_{\sum_{i=1}^{N}(r_i -m_i+1) -(v-kw)}\,
\prod_{i=1}^{N} y_i^{(r_i-m_i+1)}S_{r_i} \,.
\end{multline}
Due to the closed contour integral, with poles at $2is+v=0$, this
combination is holomorphic, and corresponds to the contribution of right-moving ground
states. The remaining $2^{N}-1$ terms make up the remainder term and
will be dealt with in the next subsection. The contour integral can be
done by picking up the poles. We have contributions whenever $v$ is an integer
that lies between $0$ and $-k+1$:
\begin{multline}
\chi_{ N,hol}(\tau,\a,\b_i)=\frac{(-1)^{N}}{\eta^{3N}(\tau)}\sum_{m_i,r_i,w}\sum_{v=-k+1}^{0} 
(-1)^{\sum_i m_i}q^{\frac{1}{2}\sum (m_i-\frac{1}{2})^2} \, z^{\sum_i (m_i-\frac{1}{2})}\cr
q^{kw^2-vw}z^{N(2w-\frac{v}{k})} \,  \delta_{\sum(r_i -m_i+1) -(v-kw)} \,
\prod_{i=1}^{N} y_i^{(r_i-m_i+1)} S_{r_i}\,.
\end{multline}
Now define the variable 
\be
p_i = r_i - m_i+1 \,,
\ee
an redefine the variable $m_i$ to $-m_i+1$ to obtain:
\begin{multline}
\chi_{ N,hol}(\tau,\a,\b_i)=\frac{1}{\eta^{3N}(\tau)}\sum_{m_i,p_i,w}\sum_{v=-k+1}^{0} 
(-1)^{\sum_i m_i}q^{\frac{1}{2}\sum (m_i-\frac{1}{2})^2} \, z^{-\sum_i (m_i-\frac{1}{2})}\cr
q^{kw^2-vw}z^{N(2w-\frac{v}{k})}\,  \delta_{\sum p_i-(v-kw)}\, 
\prod_{i=1}^{N} y_i^{p_i} S_{-m_i+p_i} \,.
\end{multline}
We now use repeatedly the formula (proven in appendix \ref{idproof}): 
\be\label{thetafrac}
\frac{i \theta_{11}(\tau,\a)}{1-zq^{p}} = \sum_{m \in \mathbb{Z}}(-1)^m q^{\frac{1}{2}(m-\frac{1}{2})^2}z^{m-\frac{1}{2}}S_{-m+p}
\ee
and we see that the holomorphic piece can be written in the rather simple form
\be\label{chiholfinal}
\chi_{ N,hol}(\tau,\a,\b_i)= \left(\frac{i \theta_{11}(\tau,-\a)}{\eta^{3}(\tau)}\right)^N\sum_{p_i,w}\sum_{v=-k+1}^{0} 
q^{kw^2-vw}z^{N(2w-\frac{v}{k})}\,\delta_{\sum p_i-(v-kw)} \, \prod_{i}\frac{y_i^{p_i}}{1-z^{-1}q^{p_i}}\,.
\ee
We thus have that a purely holomorphic piece arises from the difference of contour integrals lying on the real and the shifted real axis. 

Another form of the holomorphic state sum will be useful later on.  We
obtain it as follows.  We first flip the sign of $v$ and $w$ in the
equation \eqref{chiholfinal} above. We split the $v$-summation into
$N$ pieces, each of which goes from 
$0$ to $M-1$. We put
$v=V+jM$. Then, we obtain
\begin{multline}\label{chiNholorewrite}
\chi_{ N,hol}(\tau,\a,\b_i)= \left(\frac{i \theta_{11}(\tau,-\a)}{\eta^{3}(\tau)}\right)^N\sum_{p_i,w}\sum_{V=0}^{M-1} 
q^{kw^2-Vw}z^{-N(2w-\frac{V}{k})} \cr
\times\sum_{j=0}^{N-1}q^{-jMw}z^{j}\,\delta_{\sum p_i+V-kw+jM} \, \prod_{i}\frac{y_i^{p_i}}{1-z^{-1}q^{p_i}}\,.
\end{multline}

\subsection{Further analysis of the holomorphic part}

In this subsection we analyze the character decomposition of
the holomorphic part of the partition function, present a contour integral
representation, and argue that the Witten index equals one.

\subsubsection{Character decomposition}
We wish to read the holomorphic part of the elliptic genus as a sum
over characters. In order to facilitate this, we write the argument of
the delta function as follows:
\be
\sum_{i=1}^N p_i -v + NMw = \sum_{i=1}^N (p_i +Mw) -v
\ee
Now we define $r_i = p_i+Mw$ such
that we can write the holomorphic part as
\be\label{chiholorewriteagain}
\chi_{ N,hol}(\tau,\a,\b_i)= \left(\frac{i \theta_{11}(\tau,-\a)}{\eta^{3}(\tau)}\right)^N\sum_{r_i,w}\sum_{v=-k+1}^{0} 
q^{kw^2-vw}z^{N(2w-\frac{v}{k})}\,\delta_{\sum r_i-v}\,\prod_{i} \frac{y_i^{r_i-Mw}}{1-z^{-1}q^{r_i-Mw}} \,.
\ee
%
%
Let's proceed to exhibit the state space sum.
To connect the holomorphic contribution in equation \eqref{chiholorewriteagain}
to Ramond sector characters, we will reason in terms of a direct sum
of $N$ ${\cal N}=2$ superconformal algebras with central charge $c_f =
3 (1 + 2 N/k)=3 (1+2/M)$. In such a factor model, we start out with a
ground state character at $R$-charge $Q_R$:
\be
\chi (Q_R; q,z) = z^{Q_R-\frac{1}{2}} \frac{1}{1-z^{-1}} \frac{i \theta_{11}(\tau,-\a)}{\eta^3(\tau)} \, .
\ee
If we spectrally flow the ground state representation by $-r$ units we find:
\be
\chi (Q_R;-r; q,z)= q^{(\frac{c_f}{6}-\frac{1}{2}) r^2} z^{-(\frac{c_f}{3}-1) r} 
z^{Q_R- \frac{1}{2}} q^{(-Q_R+\frac{1}{2}) r} \frac{1}{1-z^{-1} q^r}  \frac{i \theta_{11}(\tau,-\a )}{\eta^3(\tau)} \, .
\ee
We conclude that if we flow by $-r= (-Q_R+1/2) M$ units the character of the ensuing representation reads:
\be
\chi (Q_R;-(Q_R-1/2)M; q,z) =  z^{ -r N/k }
\frac{1}{1-z^{-1} q^{r}} \frac{i \theta_{11}(\tau,-\a )}{\eta^3(\tau)}
\, .
\ee
We now consider a tensor product representation of the direct sum of the ${\cal N}=2$ superconformal
algebras. The direct sum is an ${\cal N}=2$ superconformal algebra with central charge $c=3N(1+2/M)$.
It has a character which is the product of the characters of the factor modules.
Thus, if we multiply $N$ of the characters we just constructed, with spectral flow quantum
numbers $r_i$ we find:
\be
\chi_{\otimes} (r_i;q,z) =\left(\frac{i \theta_{11}(\tau,-\a )}{\eta^3(\tau)}\right)^N  z^{-\sum_i r_i N/k } \prod_{i=1}^N
\frac{1}{1-z^{-1} q^{r_i}}   \, .
\ee
Suppose now that we only allow for representations that have a sum of individual spectral
flow quantum numbers (and therefore $R$-charges) which is equal to $v$, namely,
$\sum_{i=1}^N r_i=v$. We sum over $v$ from $0$ to $-k+1$. We then find the characters:
\be
\chi_{\otimes, \oplus} (v,r_i;q,z) = \sum_{v=-k+1}^{0}  \left(\frac{i \theta_{11}(\tau,-\a )}{\eta^3(\tau)}\right)^N z^{ -N v/k } \, \delta_{\sum r_i - v}\, 
\prod_{i=1}^N
\frac{  y_i^{r_i}}{1-z^{-1} q^{r_i}}  \, .
\ee
We have also dressed the characters with global $U(1)$ charges associated to each of the
factor models.
There is one more step to perform in order to obtain the characters of the modules featuring
in the holomorphic contribution to our path integral real Jacobi form. In the direct sum
${\cal N}=2$ superconformal algebra, we further spectrally flow by $w M$ units. Taking into account that the
relevant charge is now the total central charge, and keeping track of 
the exponents
 of $q$ and $z$ carefully,
one then obtains the characters:
\begin{align}
\chi_{fin} (w,v,r_i;q,z) &= 
\sum_{w \in \mathbb{Z}} q^{kw^2-vw} z^{2 Nw}
\sum_{v=-k+1}^{0}  \left(\frac{i \theta_{11}(\tau,-\a )}{\eta^3(\tau)}\right)^N z^{ -N v/k } 
\, \delta_{\sum r_i - v}\, \prod_{i=1}^N
\frac{  y_i^{r_i-Mw}}{1-z q^{r_i-Mw}}  \,.
\end{align}
Note that under spectral flow, the angular momentum of the state changes, since the R-charge
not only depends on fermion number but also on the angular momentum as can be seen 
by inspecting 
the exact $N=2$ superconformal algebra generators.
Summing over all $r_i$ subject to the delta function constraint we
note that this is equal to the holomorphic piece written in equation
\eqref{chiholorewriteagain}. We have thus
established the full character decomposition of the holomorphic piece.

\subsubsection{A contour integral representation}

So far we have written out the holomorphic piece of the elliptic genus
as a constrained sum over characters. It is also possible to write it
as a contour integral; the final expression is similar to the integral
expressions for the Appell-Lerch sums 
\cite{Semikhatov:2003uc}. We begin by writing the delta function in equation
\eqref{chiholfinal}
 as
\be
\delta_{\sum p_i-(v-kw)} =\frac{1}{2\pi i}\oint \frac{dx}{x} x^{\sum p_i-(v-kw)}\,.
\ee
This unconstrains the $p_i$ variables which, in turn, allows us to do the $p_i$ summation. We obtain
\be
\chi_{ N,hol}(\tau,\a,\b_i)=\frac{1}{2\pi i}\oint \frac{dx}{x}\left(\frac{i \theta_{11}(\tau,-\a)}{\eta^{3}(\tau)}\right)^N\sum_{p_i,w}\sum_{v=-k+1}^{0} 
q^{kw^2-vw}z^{N(2w-\frac{v}{k})}x^{kw-v} \prod_{i}\frac{(x y_i)^{p_i}}{1-z^{-1}q^{p_i}}\,.
\ee
Now, each of the $p_i$ summations can be done using the formula
\cite{Kac:1994kn}
\be\label{Kacformula}
\sum_{p}\frac{x^p}{1-zq^p} = \frac{i \theta_{11}(\tau,\a+\gamma)\eta^3(\tau)}{i \theta_{11}(\tau,\gamma) i\theta_{11}(\tau,\a)}\,.
\ee
Here, $x=e^{2\pi i \gamma}$ and $z=e^{2\pi i \a}$. Using this identity in the formula for the holomorphic part we get
\be
\chi_{ N,hol}(\tau,\a,\b_i)= 
\frac{1}{2 \pi i} \sum_{w \in \mathbb{Z}}\sum_{v=-k+1}^{0} \oint \frac{dx}{x} q^{kw^2-vw}z^{N(2w-\frac{v}{k})}x^{kw-v} 
\prod_{i} \left[ \frac{\theta_{11}(\tau, \gamma+ \b_i-\a)}{\theta_{11}(\tau, \gamma+ \b_i)} \right]
\,.
\ee
We also use the definition of the level $k$ theta function 
\be\label{levelktheta}
\Theta_{k,v}(\tau,\a)=\sum_{j\in \mathbb{Z}+\frac{v}{2k}}q^{kj^2}z^{kj}
\ee
to finally write the holomorphic part as the contour integral:
\be
\chi_{ N,hol}(\tau,\a,\b_i)=  
\frac{1}{2 \pi i} \oint \frac{dx}{x}\sum_{v=0}^{k-1} q^{-\frac{v^2}{4k}}x^{\frac{v}{2}}\, \Theta_{k,v}(\tau, \frac{2N\a}{k}+ \gamma)\prod_{i} \left[ \frac{\theta_{11}(\tau, \gamma+ \b_i - \a)}{\theta_{11}(\tau, \gamma+ \b_i)} \right]
\,.
\ee
This is close to expressions for truly holomorphic elliptic genera arising from gauged linear sigma models
as well as ordinary two-dimensional gauge theories.
As such it seemingly allows for an interpretation in terms of charged fields, a zero mode, and a $U(1)$
gauge field. While this is true, it
is not the whole story since modular covariance necessitates a completion.

\subsubsection{The Witten index}
To compute the Witten index of the model, it is easiest to take the
$\alpha=0$ path integral expression in equation \eqref{Nguess}. A short
calculation then gives a Witten index equal to one for all values of $N$ and
$k$.

\subsection{The modular completion of the mock modular form}
In this subsection, we return to analyzing the remainder term of the
path integral expression, namely the modular completion of the mock
modular form. We wish to massage 
this term into a compact form, and then
interpret it.

\subsubsection{A succinct expression}

Recall that we obtained the holomorphic piece by combining one term
out of $2^N$ from \eqref{chiNIIfinal} with the expression in
\eqref{firstpiece}. We are therefore left with multiple
non-holomorphic pieces:
\begin{multline}
\chi_{ N}^{rem}(\tau,\a,\b_i) =\frac{1}{\pi\eta^{3N}(\tau)}\sum_{m_i,r_i,v,w}\int_{\mathbb{R}+i\epsilon-\frac{ik}{2}} \frac{ds}{2is+v} 
(-1)^{\sum_i m_i}q^{\frac{1}{2}\sum (m_i-\frac{1}{2})^2} \, z^{\sum_i (m_i-\frac{1}{2})}q^{kw^2-vw}z^{N(2w-\frac{v}{k})} \cr
(q\bar{q})^{\frac{v^2}{4k} + \frac{s^2}{k}+is+\frac{v}{2}} 
q^{\sum_{i}(1-m_i)-v+kw} \delta_{\sum_i(r_i-m_i+1)-(v-kw)} \cr
\left[\prod_{i=1}^{N} y_i^{(r_i-m_i+1)}S_{r_i-1} - (-1)^N \prod_{i=1}^{N}y_i^{(r_i-m_i+1)}S_{-r_i}\right]\,.
\end{multline}
Here we have written the remainder as the difference of the equation
\eqref{chiNIIfinal} and the piece we take out from it, namely
\eqref{chiNIISS}. The contour is now at the shifted location in the
$s$-plane. We undo the shifts in order to
write the integral over the real axis with real exponent:
\be
s\rightarrow s-\frac{ik}{2}\qquad v\rightarrow v-k \qquad w \rightarrow w-1 \,.
\ee
%
Again, the combinations $2is+v$ and $v-kw$ are left invariant under this shift. Also, note that the imaginary part in the $(q\bar{q})$ exponent vanishes. 
Defining $n_i = m_i-1$, one can write this as 
\begin{multline}
\chi_{ N}^{rem}(\tau,\a,\b_i) =\frac{(-1)^N}{\pi\eta^{3N}(\tau)}\sum_{n_i,r_i,v,w}\int_{\mathbb{R}+i\epsilon} \frac{ds}{2is+v} 
(-1)^{\sum_i n_i}q^{\frac{1}{2}\sum (n_i-\frac{1}{2})^2} \, z^{\sum_i (n_i-\frac{1}{2})}
q^{kw^2-vw}z^{N(2w-\frac{v}{k})} \cr
(q\bar{q})^{\frac{v^2}{4k} + \frac{s^2}{k}} \, \delta_{\sum_i(r_i-n_i) -(v-kw)}
\left[\prod_{i=1}^{N} y_i^{(r_i-n_i)}S_{r_i-1} - (-1)^N \prod_{i=1}^{N}y_i^{(r_i-n_i)}S_{-r_i}\right]\,.
\end{multline}
As for the holomorphic part, we redefine the summation variables by
 setting
$p_i = r_i - n_i$
and 
we once again use the identity \eqref{thetafrac} to simplify the resulting expression to obtain
\begin{multline}
\chi^{rem}_{ N}(\tau,\a,\b_i) = \frac{1}{\pi\eta^{3N}(\tau)}\sum_{v,w}q^{kw^2-vw}z^{N(2w-\frac{v}{k})} 
\int_{\mathbb{R}+i\epsilon} \frac{ds}{2is+v} (q\bar{q})^{\frac{v^2}{4k} + \frac{s^2}{k}} \cr
\sum_{p_i \in \mathbb{Z}}\, \delta_{\sum_ip_i -v+kw}
\left[\prod_{i=1}^{N} \frac{i\theta_{11}(\tau,-\a)y_i^{p_i}}{1-z^{-1}q^{p_i}}-  \prod_{i=1}^{N}\frac{i\theta_{11}(\tau,\a)y_i^{p_i}}{1-zq^{-p_i}}\right].
\end{multline}
Using the symmetry property of the theta function and the delta
function constraint, this can be simplified to
\begin{multline}\label{Nremainder1} 
\chi^{rem}_{ N}(\tau,\a,\b_i) = \left(\frac{i\theta_{11}(\tau,-\a)}{\pi\eta^{3}(\tau)}\right)^N\sum_{v,w}q^{kw^2-vw}z^{N(2w-\frac{v}{k})} (1-z^{-N}q^{v-kw})\cr
\int_{\mathbb{R}+i\epsilon} \frac{ds}{2is+v} (q\bar{q})^{\frac{v^2}{4k} + \frac{s^2}{k}}\sum_{p_i \in \mathbb{Z}} \, \delta_{\sum_ip_i -v+kw}\prod_{i=1}^{N}\frac{y_i^{p_i}}{ (1-z^{-1}q^{p_i})}
 \,.
\end{multline}
An alternate way to write the remainder term 
is in terms of the variables $n$ and $w$ by writing $v=n+kw$. The exponent of $q$ can be simplified by completing the square and we find the following expression for the remainder:
\begin{multline}\label{Nremainder}
\chi^{rem}_{ N}(\tau,\a,\b_i) = \left(\frac{i\theta_{11}(\tau,-\a)}{\pi\eta^{3}(\tau)}\right)^N\sum_{n,w}
z^{N\frac{(kw-n)}{k}} (1-z^{-N}q^{n})\cr
\times\int_{\mathbb{R}+i\epsilon} \frac{ds}{2is+n+kw}q^{\frac{(n-kw)^2}{4k}+ \frac{s^2}{k}}\bar{q}^{\frac{(n+kw)^2}{4k} + \frac{s^2}{k}}
 \sum_{p_i \in \mathbb{Z}} \, \delta_{\sum_ip_i -n}\prod_{i=1}^{N}\frac{y_i^{p_i}}{ (1-z^{-1}q^{p_i})}
 \,.
\end{multline}

\subsubsection*{Recovering the cigar answer}

For the $N=1$ case we should recover the known cigar answer; for the holomorphic piece we obtain
\be
\chi_{ 1,hol}(\tau,\a,\b_i)= \frac{i \theta_{11}(\tau,-\a)}{\eta^{3}(\tau)}\sum_{w \in \mathbb{Z}}\sum_{v=-k+1}^{0} \, 
q^{kw^2-vw}z^{2w-\frac{v}{k}} \frac{y^{v-kw}}{1-z^{-1}q^{v-kw}} \,.
\ee
Here we have explicitly solved for the 
$p_1$ variable using the delta function. 
Similarly, setting $N=1$ in the remainder piece, we find:
\be\label{cigarrem}
\chi^{rem}_{ 1}(\tau,\a,\b_i) = 
\frac{i\theta_{11}(\tau,-\a)}{\pi\eta^{3}(\tau)}
\sum_{v,w \in \mathbb{Z}}q^{kw^2-vw}z^{2w-\frac{v}{k}}
\int_{\mathbb{R}+i\epsilon} \frac{ds}{2is+v} (q\bar{q})^{\frac{v^2}{4k} + \frac{s^2}{k}} 
y^{v-kw} \, .
\ee
We now flip the sign of $v$ and $w$ in both terms and also flip the sign of $s$ in the remainder term. 
We observe that compared to the formulas in \cite{Ashok:2011cy}, both
$z$ and $y$ appear with inverse powers. 
The charge conjugation symmetry in \eqref{cc} implies\footnote{For the $y=1$ case, this was explicitly shown in
  \cite{Ashok:2012qy} by studying the transformation properties of the
  holomorphic and remainder pieces separately. In particular, the analysis
  incorporates a careful treatment of the $\epsilon$ 
contour prescription.}
\be
\chi_{1}(\tau, -\alpha, -\b) = \chi_{1}(\tau, \a, \b)\,,
\ee
and we recover the elliptic genus obtained in \cite{Ashok:2011cy}.
\subsubsection{Observations on the modular completion}
We make a few observations on the modular completion.

\subsubsection*{The shadow}

Since we have the explicit modular completion, it is straightforward
to calculate the shadow, which is obtained as the
$\p_{\bar{\tau}}$-derivative of the elliptic genus. The only
dependence comes from the integral $I$ in the last line of
equation \eqref{Nremainder}. We obtain
\begin{align}
I &= \int_{\mathbb{R}+ i \epsilon} \frac{ds}{2is+v} (q \bar{q})^{\frac{v^2}{4k} + \frac{s^2}{k}} \, .
\cr
\p_{\bar{\tau}}I &= -\frac{i\pi}{2k}\int ds (v-2is)(q\bar{q})^{\frac{v^2}{4k} + \frac{s^2}{k}} \cr
&=-\frac{i\pi v}{4 \sqrt{k \tau_2}} (q\bar{q})^{\frac{v^2}{4k}}\,.
\end{align}
Substituting this into the $\p_{\bar{\tau}}$-derivative of the elliptic genus we obtain
\begin{multline}
\p_{\bar{\tau}}\chi_{L_N}(\tau,\a,\b_i) = -\frac{i\pi}{2}\sqrt{\frac{k}{\tau_2}}\left(\frac{i\theta_{11}(\tau,-\a)}{\pi\eta^{3}(\tau)}\right)^N\sum_{v,w \in \mathbb{Z}}q^{kw^2-vw+\frac{v^2}{4k}}z^{N(2w-\frac{v}{k})} \frac{v}{2k} \bar{q}^{\frac{v^2}{4k}}\cr
\times (1-z^{-N}q^{v-kw})\sum_{p_i \in \mathbb{Z}}\, \delta_{\sum_ip_i -v +kw}\prod_{i=1}^{N} \frac{y_i^{p_i}}{ (1-z^{-1}q^{p_i})}\,.
\end{multline}
Writing this in terms of $n=v-kw$, we find that
\begin{multline}\label{shadowfinal}
\p_{\bar{\tau}}\chi_{L_N} (\tau,\a,\b_i)= -\frac{i\pi}{2}\sqrt{\frac{k}{\tau_2}}\left(\frac{i\theta_{11}(\tau,-\a)}{\pi\eta^{3}(\tau)}\right)^N\sum_{n,w \in \mathbb{Z}} \frac{n+kw}{2k} z^{-N\left(\frac{n-kw}{k}\right)}\, q^{\frac{(n-kw)^2}{4k}}\bar{q}^{\frac{(n+kw)^2}{4k}}\cr
\times (1-z^{-N}q^{-n})\sum_{p_i \in \mathbb{Z}}\, \delta_{\sum_ip_i -n}\prod_{i=1}^{N} \frac{y_i^{p_i}}{ (1-z^{-1}q^{p_i})}\,.
\end{multline}
%
For $N>1$
we note that the sums over $n+kw$ and $n-kw$ are coupled. 
This has the following consequence. In
the examples with $N=1$ studied in \cite{Troost:2010ud, Ashok:2011cy}, the shadow
is a modular form of given weight \cite{Zwegers,Zagier,Dabholkar:2012nd}. In
these instances it is a (finite) sum over products of level $k$ theta
functions \cite{Ashok:2012qy}. As can be seen from equation \eqref{shadowfinal} this is not the
case for $N>1$. Again, as in the example of
elliptic genera of (orbifolds of) tensor product conformal field
theories \cite{Ashok:2012qy}, we see that in physical applications the
set of mock modular forms, completions and shadows that can arise is richer
than the class that is at present under good mathemical control \cite{Dabholkar:2012nd}.

\subsubsection*{A character decomposition}

The character decomposition of the remainder term is harder to
understand.  We have many more degrees of freedom than those
represented by the overall ${\cal N}=2$ superconformal algebra and thus,
we expect an infinite sum and integral over individual representations of
the overall algebra. In contrast with the holomorphic contribution,
naively taking tensor products of factor representations does not give
rise to the characters that we find. One could identify all primary states
with respect to the  overall ${\cal N}=2$ superconformal algebra (only) in the
remainder term, but that is tedious.

Rather, we can show indirectly that the expression must permit and
${\cal N}=2$ superconformal interpretation, as follows. Under modular
S-transformation, the sum of the holomorphic and non-holomorphic terms
transforms covariantly. Moreover, we know that the characters we
identified in the holomorphic part will transform into both discrete
and integrals over continuous characters
 \cite{Miki:1989ri,Eguchi:2003ik,Israel:2004xj}. The latter
contributions must conspire to give a term of the form of the
remainder term. The remainder term must then permit an interpretation
as an (albeit complicated) sum and integral over discrete and
continuous characters.

\subsubsection*{Summary}
We thus have gained insight into the decomposition of our path
integral into a holomorphic state space sum, the term that modularly
completes the mock Jacobi form to a real Jacobi form with matrix
index, and the corresponding shadow. 
 We now turn to the proposal for the identification of the
conformal field theory whose elliptic genus we have analyzed hitherto.

\section{Asymptotically linear dilaton conformal field theory}
\label{KKL}

In this section, we wish to argue that the path integral expressions
we have obtained correspond to the elliptic genera of a family of
conformal field theories that are natural generalizations of the cigar
to higher dimensions. 
We give evidence for this in terms of a gauged linear
sigma model, a counting of bound states, as well as in terms of the asymptotic 
geometries of these models.

\subsection{Asymptotically linear dilaton target spaces}

The backgrounds we consider are non-linear sigma models with
target spaces that  are asymptotically linear dilaton theories
in $d=2N$ dimensions
\cite{Kiritsis:1993pb,Hori:2001ax,Hori:2002cd}. The background metric
and dilaton that solve the string beta function equations to 
first
order in $\alpha'$ are \cite{Kiritsis:1993pb}:
\begin{align}\label{KKLgeom}
ds_{KKL}^2&= \frac{g_N(Y)}{2}dY^2 + \frac{2}{N^2 g_N(Y)}(d\psi+N A_{FS})^2 + 2 Y ds^2_{\mathbb{CP}^{N-1}}\cr
\Phi &= -\frac{NY}{k} \,.
\end{align}
Here $Y$ is a non-compact radial coordinate; $\psi$ is a periodic
variable with period $2\pi N$ and 
the circle it parameterizes is fibered over the complex
projective space $\mathbb{CP}^{N-1}$. The connection $A_{FS}$ is the
Fubini-Study connection one-form whose differential is the K\"ahler
form on $\mathbb{CP}^{N-1}$. The background is K\"ahler and has a
$U(N)$ isometry group. More details about 
the geometry and its construction can be found
in \cite{Kiritsis:1993pb, Ashok:2013kk}. 

For the case $N=1$, the background is equivalent to the cigar geometry. In this case it is known that all
$\alpha'$ corrections can be taken into account by rewriting the model
as a gauged Wess-Zumino-Witten model $SL(2,\mathbb{R})/U(1)$
\cite{Elitzur:1991cb,Mandal:1991tz,Witten:1991yr}. For $N>1$, a direct
exact conformal field theory description of the conformal fixed point
is unknown, but certain properties of the infrared theory
have been derived. First of all, we
know a gauged linear sigma model description in the ultraviolet whose
infrared fixed point 
corresponds to the non-linear sigma model described above
\cite{Hori:2001ax,Hori:2002cd}. Secondly, this ultraviolet description allows for the exact calculation of
the conformal field theory central charge: \be\label{cofKKL}
c=3N\left(1+\frac{2N}{k}\right)\,.  \ee
This precisely coincides with the central charge we have obtained by studying the modular properties of the Jacobi forms in equation \eqref{Nguess}. 

We note that it is an outstanding problem to derive the
elliptic genera of these conformal field theories from first
principles. A step in this direction was to obtain the spectrum of
particular fundamental string bound states in these higher dimensional
backgrounds. These states are labeled by (asymptotic) momentum and
winding  along the $\psi$ direction. In \cite{Ashok:2013kk} their
degeneracy was calculated by mapping the problem of finding wound
bound states to counting the ground states of a supersymmetric quantum
mechanics obtained by Scherk-Schwarz reduction of the sigma model
action along the $\psi$ circle. 

In what follows, we provide some evidence that the path integral expressions we have obtained correspond to the elliptic genera of this family of conformal field theories, labeled by
the complex dimension $N$ and asymptotic radius $R=\sqrt{k \alpha'}$.
To that end, we first argue that the gauged linear sigma model description has the salient features to give rise to our path integral result. Secondly, we show that the counting of wound bound states is
incorporated in our elliptic genus.
 Thirdly, we will argue that the remainder term is consistent with the asymptotic geometry of our models.

\subsection{The gauged linear sigma model path integral}

In this section we give heuristic arguments in support of our proposal by recalling that
there is a gauged linear sigma model (GLSM) description of the
backgrounds in equation \eqref{KKLgeom} \cite{Hori:2001ax, Hori:2002cd}.
In this description we consider a ${\cal N}=(2,2)$ supersymmetric
$U(1)$ gauge theory in two dimensions with $N$ chiral superfields
$\Phi_i$, a $U(1)$ vector multiplet $V$ and a St\"uckelberg
superfield $P$ whose imaginary part transforms additively under gauge
transformations. 
The superspace action for the gauged linear sigma model is
given by \cite{Hori:2001ax}:
\begin{equation}
\label{HKLag}
S = \frac{1}{2\pi}\int d^2x d^4\theta \left[\sum_{i=1}^N\bar{\Phi_i}e^V\Phi_i + \frac{k}{4}(P+\bar{P}+V)^2 -\frac{1}{2e^2}\bar{\Sigma}\Sigma \right] \,.
\end{equation}
Here, $\Sigma$ is a twisted chiral superfield derived from the vector multiplet.
The fields $\Phi_i$ carry charge $1$ under the $U(1)$ gauge group. 
The fermionic components of the
chiral superfields carry unit $R$-charge under the classical
vector and axial $R$-symmetries of the ${\cal N}=(2,2)$ supersymmetric
Lagrangian. The imaginary part of the $P$-field is a compact boson and
its zero mode acquires $R$-charge via the chiral anomaly \cite{Hori:2001ax}. 

In \cite{Hori:2001ax, Hori:2002cd} it is shown in detail that the gauged linear
sigma model described by the action \eqref{HKLag} flows in the infrared to a non-linear sigma-model with
the target geometry \eqref{KKLgeom}. For the case of a single chiral
superfield $\Phi$, if we write the bosonic component $\phi = \rho
e^{i\theta}$, it can be shown that the gauge invariant combination
$\text{Im}(P) -\theta$ is identified with the coordinate $\psi$ in the
cigar geometry. Similarly, $\rho$ is identified (up to a coordinate
transformation), with the radial coordiate $Y$ of the cigar.
 For the case
of multiple chiral fields $\Phi_i$, the sum of the phases of the individual
fields and the imaginary part of $P$ can be identified with the angle $\psi$,
while the other directions will make up the $\mathbb{CP}^{N-1}$ and the radial
direction.

We will now try to use this linear sigma model description to motivate
our path integral expression  \eqref{Nguess}. As explained in the
introduction, the elliptic genus of the cigar has been derived from
first principles by using a gauged Wess-Zumino-Witten description
\cite{Ashok:2011cy}. We reproduce the expression for convenience:
\be
 \chi_{cos}(\tau,\a) = k \int_0^1 ds_1
ds_2 \sum_{m,w}\left[\frac{\theta_{11}(\tau,s_1\tau+s_2-\a)}{
\theta_{11}(\tau,s_1\tau+s_2)}
e^{2\pi i \a w}\right]\, e^{-\frac{\pi k}{\tau_2}|m+w \tau +(s_1\tau+s_2+\frac{\a}{k}
  )|^2}\,.  
\ee
Here we have shifted the variable $u=s_1\tau + s_2$ by $\frac{\a}{k}$
in order to write the above expression\footnote{We are grateful to
  Nima Doroud and Jaume Gomis for clarifying this point.}.  We now
heuristically identify the ratio of theta functions appearing in the
above expression along with the phase factor as coming from the single
chiral multiplet and the non-holomorphic piece as arising from the 
(charged) compact zero mode of the imaginary part of the $P$-field.

A guiding principle in writing down the expressions for the elliptic
genera for general $N$ is  the gauged linear sigma model
description of these models. Indeed, we  include $N$ theta-function
 ratios for
the $N$ chiral superfields $\Phi_i$ in the model \cite{Witten:1993jg}, and recall that although the
action for the superfield $P$ is unchanged, the appropriate $R$-charge
of the $\text{Im}(P)$-field in the infrared is now multiplied by a
factor of $N$ \cite{Hori:2001ax}. We then see that these are the
appropriate ingredients to give rise to our path integral expression
in \eqref{Nguess}, where we have shifted the $u$ variable by
$\frac{N\a}{k}$:
\be\label{Nguessagain}
\chi_{{N}}(\tau,\a,\b_i) = k
\int_{0}^1 ds_1 ds_2 \sum_{m,w} \prod_{i=1}^N \left[\frac{\theta_{11}(\tau,s_1\tau+s_2-\a-\frac{N\a}{k}+\beta_i)}{\theta_{11}(\tau,s_1\tau+s_2-\frac{N\a}{k}+\beta_i)}\right]\, 
e^{2\pi i N \a w}\, 
e^{-\frac{\pi k}{\tau_2}|m+w \tau +(s_1\tau+s_2 )|^2}\, .
\ee
Here we have introduced the chemical potentials $\beta_i$
associated to the phase rotations of the fields $\Phi_i$, in order to match
the general proposal in \eqref{Nguess}. 
Our arguments have been
heuristic and a detailed analysis of the localization mechanism is
necessary since our model falls outside the class of models studied
for instance in \cite{Benini:2013xpa, Gadde:2013dda,
  Haghighat:2013gba}. This is because, as shown in \cite{Hori:2001ax},
the action of the $P$-field in the gauged linear sigma model is not
$Q$-exact. Consequently it is necessary to redo the localization
analysis in the gauged linear sigma-model in the presence of the
St\"uckelberg superfield $P$.\footnote{Work in progress with Nima
  Doroud.}

\subsection{The count of wound bound states}
We have given some 
arguments in support of our path integral
expression for the elliptic genus. We now provide more detailed
evidence, by comparing contributions to our proposed 
elliptic genus with contributions from wound bound states
identified in \cite{Ashok:2013kk}.

\subsubsection{Counting bound states from supersymmetric quantum mechanics}

Let us briefly review the index result we obtained in
\cite{Ashok:2013kk}, in terms of the variables used in this paper. In
that work, we started from a supersymmetric sigma model in $1+1$
dimensions with target space given by equation \eqref{KKLgeom} and
derived a supersymmetric quantum mechanical model in $0+1$ dimensions
by Scherk-Schwarz reduction of the sigma model on \eqref{KKLgeom}
along the $\psi$ direction. Effectively it gave rise
to a supersymmetric quantum mechanics with a gauge field
given by
\be\label{gauge}
A = \frac{2w}{Ng_N(Y)}(d\psi+NA_{FS}) \,.
\ee
The string winds around the $\psi$
circle $w$ times. We then calculated the (signed) number of zero mode solutions to the
Dirac equation in the background \eqref{KKLgeom}, supplemented
with the
gauge field in \eqref{gauge}. The answer to this problem can be summarized as
follows (e.g. for even $N$)\cite{Ashok:2013kk}:\footnote{Compared to
\cite{Ashok:2013kk}, we normalized $n$ such that it is
an integer, flipped its sign,  used the fact that $D(-n-N/2+1,N)=(-1)^{N-1} D(n-N/2+1,N)$
as well as the property that the overall sign of $Z_N'$ is a matter of convention, and picked
a particular regularization (discussed in \cite{Ashok:2013kk}).}
\begin{align}
Z_N'(y_1,y_2)
&=  \left[\sum_{w<0} \sum_{n=N/2}^{-kw} - \sum_{w >0} \sum_{n=-kw+1}^{-N/2}\right]
D(n-N/2+1,N) y_1^n y_2^w
\end{align}
where 
\begin{align}
D(n-N/2+1,N) &= \begin{pmatrix}
n+N/2-1\cr
N-1 
\end{pmatrix} 
\end{align}
The degeneracy factor depends on the momentum $n$ and is
independent of the winding number $w$. 
It arises from the degeneracy of Landau levels in a  quantum Hall system
on the $\mathbb{CP}^{N-1}$ section, as described in \cite{Ashok:2013kk}.
Here, $y_1$ and $y_2$ are dummy
variables that are chemical potentials for momentum and winding
respectively.  We wish to rewrite this result as a double
signed sum over wedges. To that end, we wish to perform the sum over
$n$ first. We write:
\begin{align}
Z_N'(y_1,y_2) & =  \left[\sum_{n \ge N/2} \sum_{n+kw \le 0} - \sum_{n \le -N/2 } \sum_{n+kw>0}\right]
D(n-N/2+1,N) y_1^n y_2^w \, .
\end{align}
%
The answer above is obtained by counting zero modes of a space-time fermion; consequently it is a calculation in e.g. the $NS$-$R$ sector on the 
worldsheet. In order to compare with our elliptic genus calculation, we
have to write the answer in the $R$-$R$ sector in which we worked up to now. 
Left-moving spectral  
flow will shift the
momentum  $n$  by $\frac{N}{2}$ and will leave the right-moving momentum
invariant. We then obtain:
\begin{align}
Z_N(y_1,y_2) & =  \left[\sum_{n \ge 0} \sum_{n+kw \le 0} - \sum_{n \le -N} \sum_{n+kw>0}\right]
D(n+1,N) y_1^n y_2^w \, .
\end{align}
To supply suitable weights that depend on the $q$ and $z$ variables keeping track
of conformal dimension and left-moving $R$-charge, we perform a few calculations.
Since the holomorphic contribution to the elliptic genus arises from right moving ground
states, the modes we look for have $\bar{L}_0=\frac{c}{24}$; thus, the
exponent of $q$ is equal to
\be
L_0-\frac{c}{24} = L_0- \bar{L}_0 \,.
\ee
The difference $L_0-\bar{L}_0$ is equal to the central charge of the
super quantum mechanics; this, in turn, can be obtained by acting with
the differential operator (see e.g. \cite{Stern:2000ie})
\be
Z = -iK^{\mu}\nabla_{\mu}-\frac{i}{2}(\nabla_{\mu}K_{\nu})\Gamma^{\mu}\Gamma^{\nu}
\ee
on the explicit solutions obtained in \cite{Ashok:2013kk}. 
The differential
operator depends on the Killing vector $K$ dual to the gauge field $A$ as well
as on the covariant derivative $\nabla$ and gamma matrices $\Gamma$. The operator
is the Lie derivative acting on spinors. The result of the evaluation 
is $L_0-\bar{L}_0=-nw$. This is as expected.

In order to obtain the exponent of $z$, it is necessary to know the
R-current at the IR fixed point. 
As our guideline, we will use 
the ${\cal N}=2$ superconformal algebra in terms of the
fields of the gauged linear sigma model \cite{Hori:2001ax}. We are only concerned with the
contribution to the R-charge from the momentum and winding modes on
the asymptotic circle direction. As discussed earlier, in the gauged linear sigma model,
this circle direction is identified with the imaginary part of the $P$
superfield. The contribution to the R-charge from the $Im(P)$ field
can be read off from section $7$ of \cite{Hori:2001ax} to
be\footnote{It is important to recall that the $P$ field is not
  canonically normalized in the gauged linear sigma model action of \cite{Hori:2001ax}.}
\be
N\left(\frac{n-kw}{k}\right)\,.
\ee
We believe this answer is also valid at the infrared fixed point.
Putting these facts together we find the following expression as a
contribution to the elliptic genus:
\begin{align}\label{SQMresult}
Z_N(q,z) & =  \left[\sum_{n \ge 0} \sum_{n+kw \le 0} - \sum_{n \le -N} \sum_{n+kw>0}\right]
D(n+1,N) q^{-nw} z^{\frac{N(n-kw)}{k}} \, .
\end{align}

\subsubsection{Matching wound bound states}

In order to obtain the wound bound states of fundamental strings in
the geometry (\ref{KKLgeom}), we wound a string on the asymptotic
$\psi$-circle, and computed the number of bound states, given that
neither oscillator excitations nor other winding numbers were turned
on. 
We saw that these bound states carried left-moving conformal dimension equal
to 
$-nw+c/24$, which fixes the power of the modular parameter $q$ in the
partition sum to be of the product form.  We now identify the
particular terms in the holomorphic part of the elliptic genus corresponding
to these particular bound states. 
Recall that the elliptic genus is charge conjugation symmetric, as noted
in equation (\ref{cc}). For easier comparison to the index result of  \cite{Ashok:2013kk},
we will work with the charge conjugate holomorpic contribution:
\be
\chi_{ N,hol}(\tau,\a,\b_i)= \left(\frac{i \theta_{11}(\tau,\a)}{\eta^{3}(\tau)}\right)^N\sum_{p_i,w}\sum_{v=-k+1}^{0} 
q^{kw^2-vw}z^{-N(2w-\frac{v}{k})}\,\delta_{\sum p_i-v+kw} \, \prod_{i}\frac{y_i^{p_i}}{1-zq^{p_i}}\,.
\ee
To identify the relevant states in our partition sum, we recall the
identification of the quantum number $v$ with the right-moving
asymptotic momentum $n+kw$, 
such that the constraint reads 
\be
\sum_{i=1}^N p_i = n \, .
\ee
In order to expand the denominators in the product, we assume that
$|q|<|z|<1$. 
We further assume that all
integers $p_i$ and $n$ are of the same sign. For $n \ge
0$, we find that the number of solutions for the integers $p_i$ is
given by
\be
\begin{pmatrix}
n+N-1 \cr  
N-1
\end{pmatrix} 
\, . 
\ee
On the other hand, for $n <0$, the degeneracy is
\be
\begin{pmatrix}
-n-1 \cr  
N-1
\end{pmatrix} \, ,  \ee 
up to an overall sign.
These two expressions are related by a factor
$(-1)^{N-1}$.  The number of solutions therefore matches the degeneracy of the wound bound string
ground states. Although we already identified the degeneracy,
we still need to argue that we can freely sum over momentum and
winding, since our holomorphic partition sum only contains a sum over
the number $v$ in a particular range.
To that end, we expand our partition sum 
using the assumption that either all $p_i$ are positive,
or strictly negative. These two possibilities (out of $2^N$) lead to the terms:

\begin{multline}
Z_{hol,N}(\tau,\a,\b_i) = \sum_{w \in \mathbb{Z}} \sum_{v=-k+1}^0 z^{Nv/k} q^{kw^2-vw}
z^{- 2Nw}   \cr
\times \left[\sum_{p_i \ge 0} \sum_{w_i \ge 0} z^{\sum_i w_i} q^{\sum_i w_i p_i}
+\dots
+ (-1)^N \sum_{p_i<0} \sum_{w_i \ge 0} z^{-N-\sum_i w_i} q^{-\sum p_i} q^{ - \sum_i w_i p_i}\right] \delta_{\sum p_i = n}\prod_{i=1}^N y_i^{p_i} \, .
\end{multline}
We have dropped the prefactors corresponding to oscillator excitations not captured by the supersymmetric
quantum mechanics. We wish to think of $w_i$ as a winding number associated to the phase of the
projective coordinate $\Phi_i$ of $\mathbb{CP}^{N-1}$. The strings wound in \cite{Ashok:2013kk} wound
only the overall $\psi$ coordinate, or in other words, each $\Phi_i$ phase an equal number of times.
We therefore restrict to windings $w_i=r$ in the above sum and find:
\begin{multline}
Z_{restr, hol,N} (\tau,\a,\b_i)= \sum_{w \in \mathbb{Z}} \sum_{v=-k+1}^0 q^{-nw}
z^{- Nw}  z^{Nn/k} \cr
\left[  \sum_{p_i \ge 0} \sum_{r \ge 0} z^{Nr} q^{r n}
+\dots
+ (-1)^N \sum_{p_i<0} \sum_{r \ge 0} z^{-N(r+1)} q^{-(r+1)n} \right] \prod_{i=1}^N y_i^{p_i} \delta_{\sum p_i = n} \, .
\end{multline}
%
In the summation term with all $p_i>0$, we introduce the new variable
\be
\tilde{w} = w-r \,,
\ee
in terms of which we see that the constraints $-(k-1) \le
v \le 0$ and $n\ge 0$ as well as $r \ge 0$ map to the conditions $n+k\tilde{w} \le 0$
and $n \ge 0$.
 The summation over the finite range for  $v$ as well as the constraint $v=n+kw$
imposes a strong constraint on the sum over $w$. This constraint is removed through the shift by $r$,
and the sum over the finite range of $v$ and the infinite range of $r$ is replaced by
a summation over the integer $n+k \tilde{w}$. The sum over $n$ is then free, up to the constraint
on its sign.
Similarly in the last term, we consider the change of variables:
\be
\tilde{w} = w+r+1 \, .
\ee
For these terms, the constraints become $n+k\tilde{w} > 0$ and
$n < 0$. 
 The sum
over
the integers $p_i$ subject to the constraint precisely reproduces the
degeneracy factor, as mentioned earlier.  The  chemical
potentials $y_i$ keep track of the origin of the degeneracy in the sum over
$U(1)^N \subset U(N)$ angular momenta. If we put $y_i=1$, we find (after removing
the tilde from the variable $\tilde{w}$): 
\begin{multline}
Z_{restr, hol,N}(\tau,\a) = \left[\sum_{n \ge 0} 
\sum_{n+kw \le 0}+\ldots - \sum_{n \le -N} \sum_{n+kw > 0}  
\right]
\begin{pmatrix}
n+N-1 \cr  
N-1
\end{pmatrix}
q^{-nw} z^{N(\frac{n-kw}{k})}  \,.
\end{multline}
We have canceled the sign $(-1)^N$ against the factor $(-1)^{N-1}$ that arose when
counting the number of solutions.
We have found a precise agreement with equation \eqref{SQMresult}.

It would be interesting to generalize the supersymmetric quantum
mechanics that arose from Scherk-Schwarz reduction on the
$\psi$-circle in \cite{Ashok:2013kk} to the case where we allow for
windings along other angular coordinates.  We also expect
 these indices to be captured by our proposal for the
elliptic genus.

\subsection{Features of the asymptotic geometry and susy algebra}
Finally, we turn to arguing that the remainder term is in accord with the asymptotic
geometry of our model.
Indeed, from the form of the modular completion, one can infer several things
about the conformal field theory. For the case of the cigar (at $N=1$) \cite{Ashok:2011cy},
the remainder function in \eqref{cigarrem} can be interpreted as a sum over states in the cigar
conformal field theory. One interprets the $s$-integration as an
integral over the radial momentum of the states; in
\cite{Ashok:2011cy} it was shown that the measure of the $s$ integral
is given by the spectral asymmetry between the bosons and fermions in
the continuum sector of the conformal field theory. 
In \cite{Ashok:2013kk}
we argued the difference in density is fixed in terms of the asymptotic right-moving
supercharge. The sum over $n$
and $w$ can be interpreted as the sum over momentum and winding modes
on the asymptotic circle direction in the cigar geometry. The
combination $v=n+kw$ is the overall
right-moving momentum.

Applying the same logic to the expressions for the $N>1$ case, we
notice many similarities and a few important differences. One can read
off a few features immediately from the remainder term in
\eqref{Nremainder}: there is a single radial momentum $s$; the measure
of the integral is identical to the cigar case and is determined by a
linear combination of the radial momentum $s$ and the overall
right-moving angular momentum $v$. The asymptotic geometry of the
target space sigma model is therefore the combination of a radial
direction, an overall asymptotic circle, which is the
superpartner of the radial direction, and a compact section.
The
quantum numbers $n$ and $w$ are interpreted as the momentum and
winding along the asymptotic circle direction.  In the cigar geometry
with asymptotics $\mathbb{R}\times S^1$, the left and right moving
momenta along the circle are decoupled. The lack of decoupling for the
$N>1$ cases shows that the asymptotic circle direction is fibered
over the compact section.
All these features are in perfect agreement with the asymptotic geometry of the space
(\ref{KKLgeom}). A more detailed analysis should show that even the precise compact
section is coded in the degeneracies in the remainder term.

\subsubsection*{Summary}
We have shown that our elliptic genus includes 
the states that were counted in the twisted index calculation of
\cite{Ashok:2013kk}. Moreover, we argued that the remainder term is in accord
with the asymptotic geometry of the models.
Combined with the modular and elliptic properties
of our proposal (which gives the central charge), as well as the heuristics
based on the gauged linear sigma model, we believe that this
provides good evidence for our identification of the conformal field theory
that leads to the Jacobi forms \eqref{Nguess} as the sigma model on the asymptotic linear
dilaton background  \eqref{KKLgeom}. 

\section{Elliptic genus of generalized Liouville theories}
\label{genLiou}

The elliptic genus of Liouville theory \cite{Troost:2010ud} also
allows for a generalization to a family labeled by an extra integer
$N$.  We summarize the salient features of the proposal in this case.
Many of the technical details are omitted since the calculations
follow closely those that we have presented in sections
\ref{pathprop} and \ref{decomposition}. We begin with the
following proposal:
\begin{multline}\label{LNguess}
\chi_{L_{N}}(\tau,\a,\b_i)  = \int_0^1 ds_1 ds_2 \sum_{m,w}
\prod_{i=1}^N\left[\frac{\theta_{11}(\tau,s_1\tau+s_2-\a-\frac{N\a}{k} + \b_i)}{\theta_{11}(\tau,s_1\tau+s_2-\frac{N\a}{k}+\b_i)}\right]
\, e^{\frac{2\pi i N\a w}{k}}e^{-\frac{\pi }{k\tau_2}|m+\tau w +k(s_1\tau+s_2)|^2}\,.
\end{multline}
This is a generalization of the elliptic genus of Liouville theory
\cite{Troost:2010ud}. We have twisted by $N$
 $U(1)$ global symmetries. In order to derive the modular
and elliptic properties, we do a Poisson resummation to obtain
\begin{multline}\label{LNguesstwo}
\chi_{L_{N}}(\tau,\a,\b_i)  = k \int_0^1 ds_1 ds_2 \sum_{m,w}
\prod_{i=1}^N\left[\frac{\theta_{11}(\tau,s_1\tau+s_2-\a-\frac{N\a}{k} + \b_i)}{\theta_{11}(\tau,s_1\tau+s_2-\frac{N\a}{k}+\b_i)}\right]\cr
\times e^{-2\pi i k s_2 w}e^{2\pi i s_1(km-N\a)}e^{-\frac{\pi k}{\tau_2}|m-\frac{N\a}{k}+w\tau|^2}\,.
\end{multline}
The modular and elliptic properties can be checked to be same as those
found in equation \eqref{cigarmod-ell}, with the same central
charge. We 
therefore have another path integral expression for a real Jacobi
form with matrix index.

As was shown in
\cite{Eguchi:2010cb, Ashok:2011cy} the elliptic genus of the cigar and
Liouville models (at the same asymptotic radius $\sqrt{k\a'}$) are
related by a $\mathbb{Z}_k$ orbifold. Analogously, we will show later
in the section that the above two models are  related by a
$\mathbb{Z}_{\frac{k}{N}}$ orbifold. But first we 
obtain the holomorphic and remainder pieces of the Liouville
generalization.

\subsection{The mock modular form and its completion}

To extract the holomorphic piece from the path integral expression, the analysis
proceeds as before. We perform a single Poisson resummation on equation \eqref{LNguesstwo} to obtain

\be
\chi_{L_ N}(\tau,\a,\b_i) = \sqrt{k \tau_2} \int_0^1 ds_1 ds_2 \sum_{n,w} \prod_{i=1}^N
\left[\frac{\theta_{11}(s_1\tau+s_2-\a-\frac{N\a}{k} +\beta_i, \tau)}{
\theta_{11}(s_1\tau+s_2-\frac{N\a}{k}+\beta_i, \tau)}\right]\, 
e^{\frac{2\pi i N\a n}{k}}\, 
e^{-2\pi i k s_2 w}\,
q^{\ell_0} \bar{q}^{\bar{\ell}_0}\,,
\ee
where $\ell_0$ and $\bar{\ell}_0$ are given by
\be
\ell_0 = \frac{(n+k(s_1-w))^2}{4k}\qquad \bar{\ell_0} = \frac{(n+k(s_1+w))^2}{4k}\,.
\ee
We expand the theta functions in power series using equations
 \eqref{denomexpand} and \eqref{numexpand}; the 
holonomy integral over $s_2$  now leads to the constraint
\be
\sum_{i}(r_i-m_i+1) = kw\,.
\ee
Substituting 
the constraint into the expression gives
\begin{multline}
\chi_{L_N}(\tau,\a,\b_i)=  (-1)^N \frac{\sqrt{k \tau_2}}{\eta^{3N}(\tau)}\sum_{m_i,r_i,n,w}
\int_0^1 ds_1 (-1)^{\sum_i m_i}q^{\frac{1}{2}\sum (m_i-\frac{1}{2})^2} \, z^{\sum_i (m_i-\frac{1}{2})}\cr
q^{-nw}z^{\frac{N}{k}(n-kw)}(q\bar{q})^{\bar{\ell}_0} \delta_{\sum_i(r_i-m_i+1) -kw}  \prod_{i=1}^N y_i^{r_i-m_i+1} S_{r_i}  \,.
\end{multline}
We perform the integral over the holonomy $s_1$ by first introducing a
radial momentum $s$ (in order to linearize the $s_1$ exponent). We
also introduce the right-moving momentum variable $v=n+kw$ and follow the steps
that were followed for the models in section \ref{decomposition}. We obtain a sum over $2^N + 1$
terms, out of which precisely two terms combine to produce a contour
integral which can be performed to give the purely holomorphic part of
the elliptic genus:
\be\label{chiholLN}
\chi_{L_N,hol}(\tau,\a,\b_i)=\left(\frac{i\theta_{11}(\tau,-\a)}{\eta(\tau)^3}\right)^N 
\sum_{w \in \mathbb{Z}} q^{kw^2}z^{-2Nw} 
\sum_{v=0}^{k-1} (z^{-\frac{N}{k}}q^{w})^v
\sum_{p_i \in \mathbb{Z}}\delta_{\sum p_i - kw}\prod_{i=1}^N \frac{y_i^{p_i}} {(1-z^{-1}q^{p_i}) } \,.
\ee

\subsubsection{The completion}

The remaining $2^N-1$ terms make up
 the remainder function:
\begin{multline}\label{L_Nremainder}
\chi^{rem}_{L_N}(\tau,\a,\b_i) = \left(\frac{i\theta_{11}(\tau,\a)}{\pi\eta^{3}(\tau)}\right)^N
\sum_{w \in \mathbb{Z}} \sum_{v=-k+1}^0
q^{kw^2-vw}z^{-N(2w-\frac{v}{k})} \cr
\times\sum_{p_i \in \mathbb{Z}}\frac{(1-z^{-N}q^{kw})}{\prod_{i=1}^{N} y_i^{-p_i} (1-z^{-1}q^{p_i})}\, \delta_{\sum_ip_i -kw}
\int_{\mathbb{R}+i\epsilon} \frac{ds}{2is+v} (q\bar{q})^{\frac{v^2}{4k} + \frac{s^2}{k}} \,.
\end{multline}

\subsubsection{Integral representation}
An integral expression for the holomorphic part can be
 obtained by writing the delta function as an integral:
\be
\chi_{L_N,hol}(\tau,\a,\b_i)=\frac{1}{2 \pi i } \oint \frac{dx}{x}\left(\frac{i\theta_{11}(\tau,-\a)}{\eta(\tau)^3}\right)^N 
\sum_{w \in \mathbb{Z}} q^{kw^2}(z^{-2N}x^{-k})^w 
\sum_{v=0}^{k-1} (z^{-\frac{N}{k}}q^{w})^v
\sum_{p_i \in \mathbb{Z}}\prod_{i=1}^N \frac{(xy_i)^{p_i}} {(1-z^{-1}q^{p_i}) }
\ee
Making use of the formulas \eqref{Kacformula} and \eqref{levelktheta}, this can be written in the compact form
\be
\chi_{L_N,hol}(\tau,\a,\b_i) = \frac{1}{2 \pi i }\oint \frac{dx}{x}
\sum_{v=0}^{k-1}q^{-\frac{v^2}{4k}}x^{\frac{v}{2}}\Theta_{k,v}(\tau, -\gamma -\frac{2N\a}{k})\, \prod_{i=1}^{N}\frac{\theta_{11}(\tau, \gamma + \b_i -\a)}{\theta_{11}(\tau, \gamma+ \b_i)} \, .
\ee

\subsubsection{Poincar\'e polynomial}
The Poincar\'e polynomial of the model is given by the  $q \rightarrow 0$ limit of the Liouville elliptic
genus. The $\Theta$-function at level $k$ reduces to a single term in
this limit.
The Poincar\'e polynomial is $y_i$ independent, and equal to:
\be
\chi_{L_N,hol} (\a)
= \frac{z^{N/2} -z^{-N/2}}{1-z^{-N/k}} \, .
\ee
Thus, we have $k$ ground states, 
with R-charges distributed symmetrically around zero, and quantized in units
of $N/k$. They carry no global charge.

\subsubsection{Character interpretation}

In order to
find the character representation of the holomorphic piece, we write
the delta function constraint in equation \eqref{chiholLN} as
\be
(\sum_{i=1}^N p_i)-kw = \sum_{i=1}^N (p_i - Mw) \,,
\ee
where we have used $k=NM$. Defining $r_i = p_i - Mw$, we find that
\be\label{chiholLNsumzero}
\chi_{L_N,hol}(\tau,\a,\b_i)=\left(\frac{i\theta_{11}(\tau,-\a)}{\eta(\tau)^3}\right)^N 
\sum_{w \in \mathbb{Z}} q^{kw^2}z^{-2Nw} 
\sum_{v=0}^{k-1} (z^{-\frac{N}{k}}q^{w})^v
\sum_{r_i}\delta_{\sum r_i} \prod_{i=1}^N \frac{y_i^{r_i+Mw}} {(1-z^{-1}q^{r_i+Mw}) }
\,.
\ee

Let us now interpret 
the result in terms of a constrained sum of products of
Ramond sector characters. We begin with Ramond ground states with
R-charge $Q_R$.
%
We take the tensor product of $N$ of these representations, and spectrally flow the individual factors
by $-r_i$ units to find:
\be
\chi_{\otimes} (Q_R;-r_i;q,z) =\prod_{i=1}^N
q^{(\frac{c_f}{6}-\frac{1}{2})  r_i^2} z^{-(\frac{c_f}{3}-1) r_i} z^{Q_R-\frac{1}{2}} q^{ (-Q_R+\frac{1}{2}) r_i} 
 \frac{1}{1-z^{-1} q^{r_i}}  \left(\frac{i \theta_{11}(\tau,-\a )}{\eta^3(\tau)} \right)^N
\ee
Now assume that $-Q_R+1/2= v/k - r_i/M$ in each of the $N$ sectors. We
then see that the quadratic term in the exponent cancels. We
moreover must assume that $\sum_{i=1}^N r_i=0$. We thus find, at this stage:
\be 
\chi_{\otimes} (Q_R,r_i;q,z) = \left(\frac{i \theta_{11}(\tau, -\a)}{\eta^3(\tau)} \right)^N z^{ -\frac{N v}{k}} \delta_{\sum_{i=1}^N r_i = 0}
\prod_{i=1}^N \frac{1}{1-z^{-1} q^{r_i}} 
\ee 
We next perform spectral flow by
$-Mw$ units in the direct sum of the factor ${\cal N}=2$ superconformal
algebras. Summing over all $r_i$ subject to the delta function
constraint, we obtain expression \eqref{chiholLNsumzero}, and therefore, the character sum
interpretation of the holomorphic contribution.

\subsection{Orbifolds}

We will now show that the two classes of models we have discussed so
far are related by a $\mathbb{Z}_M$ orbifold,
where the discrete orbifold group is a subgroup of the $U(1)$
$R$-symmetry group. There is a systematic way to construct
$R$-symmetry orbifolds, following \cite{Kawai:1993jk}. Starting with
the elliptic genus, we first define the ``twisted blocks":
\be
\chi_{m_a,m_b}(\tau,\a,\b_i) = (-1)^{\frac{c}{3}m_am_b}q^{\frac{c m_a^2}{6}}z^{m_a \frac{c}{3}}
\chi(\tau, \alpha+m_a \tau + m_b)\prod_{i}y_i^{-m_a} \, .
\ee
The integers $m_a, m_b \in \mathbb{Z}_M$ label the twisted sectors in the orbifold theory.
For the case at hand we work with the holomorphic part of the Liouville elliptic genus, the discussion with the remainder can be done in an analogous manner. Using the definition above we obtain the twisted blocks (after a simplification where we shift the $p_i$ variable by $m_a$):
\begin{multline}
\chi_{L_N,m_a,m_b} (\tau,\a,\b_i)= (-1)^{N(m_a+m_b+m_a m_b)}e^{\frac{2\pi i m_b}{M}(Nm_a-v)} q^{\frac{m_a^2N}{M}} z^{\frac{2m_aN}{M}} \cr
\left(\frac{i \theta_{11}(\tau,-\a)}{\eta^{3}(\tau)}\right)^N\sum_{w \in \mathbb{Z}}\sum_{v=0}^{k-1} q^{kw^2}z^{-2Nw}(z^{-\frac{N}{k}}q^w)^v  q^{-2wNm_a-\frac{vm_a}{M}}\cr
\sum_{p_i \in \mathbb{Z}}\prod_{i=1}^N \frac{y_i^{p_i}}{1-z^{-1}q^{p_i}}\delta_{\sum p_i - kw+Nm_a} \,.
\end{multline}
It can then be shown, on general grounds (see \cite{Kawai:1993jk}), that the following sum over the twisted blocks satisfies all the requirements to be the elliptic genus of a conformal field theory with the same central charge:
\be
\chi_{L_N;\mathbb{Z}_M}(\tau,\a,\b_i) = \sum_{m_a, m_b \in \mathbb{Z_M}} (-1)^{N(m_a+m_b+m_am_b)}\chi_{L_N, m_a, m_b}(\tau,\a,\b_i) \,.
\ee
The sum over the $m_b$ variable leads to the constraint $Nm_a = v$ modulo $M$. The
solutions to this can be obtained as follows: write $v = V +jM$, where
$V=0,1,2 \ldots M-1$ and $j=0,1,2, \ldots N-1$. We have split the
$v$-summation into $N$ intervals, each of length $M$. Then, one can
check that for every value of $m_a$, there are exactly $N$ solutions
to the constraint, one in each of the $N$ sub-intervals of the
original $v$-summation. Substituting this into the sum over the
twisted blocks, 
we obtain the result
%
\begin{align}
\chi_{L_N,hol;\mathbb{Z}_M}(\tau,\a,\b_i) &= \left(\frac{i \theta_{11}(\tau,-\a)}{\eta^{3}(\tau)}\right)^N \sum_{w \in \mathbb{Z}}\sum_{V=0}^{M-1}
q^{kw^2-vw}
z^{-N(2w-\frac{V}{k})}
\sum_{j=0}^{N-1}q^{-jMw}z^{j} 
\cr
&\qquad\qquad\times  \sum_{p_i \in \mathbb{Z}} \prod_{i=1}^N \frac{y_i^{p_i}}{1-z^{-1}q^{p_i}}\delta_{\sum p_i -kw+V+jM}
\cr
& \equiv  \chi_{N,hol}(\tau,\a,\b_i) \,.
\end{align}
In the second equality we have made the identification of the orbifold
elliptic genus with the holomorphic part written out in the form of equation
\eqref{chiNholorewrite}.  A similar calculation can be performed to
show that the remainder functions are also related by a $\mathbb{Z}_M$
orbifold. We therefore find that, much like the original case of the
Liouville and cigar theories at the same asymptotic radius, the two
path integrals in \eqref{Nguess} and \eqref{LNguess} are related by an
overall $\mathbb{Z}_M$ orbifold.

\section{Conclusions}
\label{conclusions}

In this paper we have exhibited interesting real Jacobi forms with
matrix index parameterized by two integers $k$ and $N$. For a given
value of these parameters we have provided evidence that the Jacobi
form arises as the elliptic genus of a non-compact conformal field
theory with central charge $c= 3N(1+2N/k)$. There is a geometric
description of this conformal field theory as a complex
$N$-dimensional K\"ahler manifold that has an $S^1$ fibered over the
complex projective space $\mathbb{CP}^{N-1}$, along with a radial
direction which has an asymptotic linear dilaton. 
When $N$ divides $k$, we also showed that a $\mathbb{Z}_{\frac{k}{N}}$ orbifold of our
proposal gave rise to another class of elliptic genera with the same
central charge. These can be understood as multi-variable
generalizations of Liouville theories that also have a gauged linear
sigma model description \cite{Hori:2001ax}.
 Indeed, we expect that our
techniques apply to the whole zoo of models described in
\cite{Hori:2001ax,Hori:2002cd}. Moreover, the identification of the
elliptic genera of this large class of models opens a window onto
their full spectrum.

The conformal field theory backgrounds that we studied appear in
string theory when we consider the worldsheet description of $NS5$
branes wrapped on $\mathbb{CP}^{N-1}$. Thanks to our results, it has
become straightforward to calculate the worldsheet elliptic genera in
such backgrounds.  An application of these results is to compute a
space-time index that arises from the worldsheet (generalized)
elliptic genus through integration over the fundamental domain. This
potentially generates interesting mock modular forms with a direct
space-time interpretation. In
\cite{Harvey:2013mda}
 an example of this technique was exhibited,
by considering the near horizon geometry of two $NS5$ branes wrapping
a $K3$ surface. The worldsheet theory which describes such a
background includes the cigar elliptic genus; it will be interesting
to generalize these results to other models at $N=1$ as well as to
higher $N$ -- we have only seen the beginning of these
applications.

A first principles derivation of the elliptic genus is desirable.
It should be attainable
by applying localization techniques to the gauged linear sigma model
(GLSM) description of these backgrounds \cite{Hori:2001ax,
  Hori:2002cd}. This is especially interesting because there are other
non-compact backgrounds that also have 
a gauged
linear sigma model description, such as the Euclidean Taub-NUT
background to which our techniques could be adapted.  A further
challenge includes the understanding of a conjectured elliptic genus
of Atiyah-Hitchin space \cite{Haghighat:2012bm}.

Our elliptic genera exhibit shadows which (like
those of the orbifolded tensor product models of \cite{Ashok:2012qy})
suggest that the realm of mock modular forms may be usefully defined
even beyond the class that is at present mathematically well-understood
\cite{Dabholkar:2012nd}.
Producing qualitatively new examples of non-compact elliptic genera,
as we did, should be helpful in coming to grips with a grand
synthesis.

\section*{Acknowledgments}
We would like to thank Atish Dabholkar, Nima Doroud, Jaume Gomis, Dan
Israel, Bruno Le Floch, Sameer Murthy, Suresh Nampuri and Giuseppe
Policastro for useful and interesting discussions. We thank the JHEP
referee for constructive criticism.  This work was
supported in part by the ANR grant ANR-09-BLAN-0157-02.

\begin{appendix}

\section{Formulas}
\subsection{Definitions and properties}
The Jacobi theta function is given by
\be
\theta_{11}(\tau,\a) = -i\sum_{n=-\infty}^{\infty} (-1)^n q^{\frac{1}{2}(n-\frac{1}{2})^2}z^{n-\frac{1}{2}} \, , 
\ee
where $q=e^{2\pi i \tau}$ and $z=e^{2\pi i \a}$. 
The Dedekind eta function is given by
\be
\eta(\tau) = q^{\frac{1}{24}}\prod_{n=1}^{\infty}(1-q^n)\,.
\ee
The modular and elliptic properties of the combination
 $\theta_{11}(\tau,\alpha)/\eta^3(\tau)$
are:
\be
\frac{\theta_{11}}{\eta^3}\left(\frac{a\tau+b}{c\tau+d}, \frac{\a}{c\tau+d}\right) = 
(c\tau+d)^{-1}e^{\pi i\frac{c\a^2}{c\tau+d}}\frac{\theta_{11}}{\eta^3}(\tau, \a) \,.
\ee
\be
\frac{\theta_{11}}{\eta^3}(\tau, \a+m_a\tau+m_b) = (-1)^{m_a+m_b}q^{-\frac{m_a^2}{2}}z^{-m_a}\frac{\theta_{11}}{\eta^3}(\tau, \a) \, .
\ee
%
%
We also use the level $k$ theta function, defined to be

\be
\Theta_{k,v}(\tau,\a)=\sum_{j\in \mathbb{Z}+\frac{v}{2k}}q^{kj^2}z^{kj} \, .
\ee
\subsection{Identities and their proof}
\label{idproof}
In this subsection of the appendix, we list a number of identities, and their proofs.
We use the series
\be S_r(q) = \sum_{n=0}^{\infty}(-1)^nq^{\frac{n(n+2r+1)}{2}} \, , \ee
which is related to the inverse of the $\theta_{11}$ function by: \be
\frac{1}{i \theta_{11}(\tau, \a )} = \frac{1}{\eta^3(\tau)} \sum_{r \in
  \mathbb{Z}}z^{r+\frac{1}{2}}S_{r}(q) \,, \ee 
when the arguments
satisfy $|q|<|z|<1$. This identity can be proven by matching
coefficients of powers of $z$ on both sides of the identity. Indeed,
let's consider the contour integral:
\begin{eqnarray}
C(q) &=& \frac{1}{2\pi i } \oint \frac{dz}{z^{r+\frac{3}{2}}} \frac{1}{i \theta_{11}(\tau,\a)} \, ,
\end{eqnarray}
where the contour lies in the annulus $|q|<|z|<1$ and circles
the origin counter clockwise once. We use the product form of the Jacobi 
theta function to find:
\begin{eqnarray}
C(q) &=& - q^{-\frac{1}{8}} \frac{1}{2 \pi i} 
\oint \frac{dz}{z^{r+\frac{3}{2}}} \frac{1}{ (z^{\frac{1}{2}}-z^{-\frac{1}{2}})
\prod_{n=1}^\infty (1-q^n)(1-zq^n)(1-z^{-1}q^n)} \, .
\end{eqnarray}
We can compute the  integral by picking up the poles inside the contour. These lie at
$z = q^{m}$ where $m \in \{ 1,2,3, \dots \}$. We pick up the residues at these poles,
which are equal to:
\begin{eqnarray}
R_m(q) &=& \frac{ (-1)^{m-1} q^{-mr+m(m-1)/2} }{\eta^3(\tau)} \, . 
\end{eqnarray}
We then find the result of the contour integration:
\begin{eqnarray}
C(q) &=& \sum_{m=1,2,\dots}^\infty  \frac{ (-1)^{m-1} q^{-mr+m(m-1)/2} }{\eta^3(\tau)}
\nonumber \\ 
&=& \sum_{m=0,1,\dots}^\infty  \frac{ (-1)^{m} q^{-mr+m(m+1)/2} q^{-r} }{\eta^3(\tau)}
\nonumber \\
&=&  
\frac{1}{\eta^3(\tau)} q^{-r} S_{-r}(q)= \frac{1}{\eta^3(\tau)} S_r(q)  \, .
\end{eqnarray}
In the last line, we use the equality:
\begin{eqnarray}
\label{flip}
q^{-r} S_{-r}(q) &=& S_r(q) \, ,
\end{eqnarray}
which can be proven by observing that the difference of these two expressions is a finite
sum whose terms cancel two by two. We thus have proven the desired equality, in the 
particular range of arguments.  Let us also observe that the equality:
\begin{eqnarray}
S_r(q) + S_{-r-1}(q) &=& 1
\end{eqnarray}
is now easily proven by shifting the summation variable in $S_{-r-1}$ by one and using
equation (\ref{flip}). Alternatively, these identities satisfied by the series $S_r$ can be found
by using properties of the $\theta_{11}$-function, and contour integration.

Another identity we put to good use is the expansion:
\be 
\frac{i \theta_{11}(\tau,\a)}{1-zq^{p}} = \sum_{m \in \mathbb{Z}}(-1)^m q^{\frac{1}{2}(m-\frac{1}{2})^2}z^{m-\frac{1}{2}}S_{-m+p}(q) \, .
\ee
We can prove this identity, for instance for $|z q^p|<1$, by expanding the denominator and then
rearranging terms, as follows:
\begin{eqnarray}
\frac{i \theta_{11}(\tau,\a)}{1-zq^{p}} &=& \sum_{m \in \mathbb{Z}}
(-1)^m q^{ \frac{(m-\frac{1}{2})^2}{2}} z^{m - \frac{1}{2}} 
\sum_{l=0,1,\dots}^{\infty} (z q^p)^l
\nonumber \\
&=& \sum_{m=-\infty}^{+\infty}
(-1)^m q^{ \frac{(m-\frac{1}{2})^2}{2}} z^{m - \frac{1}{2}} 
\sum_{l=0,1,\dots}^{\infty}  q^{(p-m)l + l (l+1)/2}
\nonumber \\
&=& \sum_{m=-\infty}^{+\infty}
(-1)^m q^{ \frac{(m-\frac{1}{2})^2}{2}} z^{m - \frac{1}{2}} S_{-m+p}(q) \, .
\end{eqnarray}
In going from the first to the second line, we have shifted the summation variable
$m$ by $-l$. For $|z q^p|>1$, the identity can be proven analogously.

\end{appendix}

\end{document}